\documentclass[pra,aps,twocolumn,floatfix,showpacs,10pt]{revtex4-1}
\usepackage{amsmath}
\usepackage{graphicx}
\usepackage{epsfig}
\usepackage{graphics}
\usepackage{bm}
\usepackage{bbm}
\usepackage{amssymb}
\usepackage{psfrag}
\usepackage{mathtools}
\usepackage[english]{babel}
\usepackage[T1]{fontenc}
\usepackage{booktabs}
\usepackage{color}
\usepackage{hyperref}
\usepackage{braket}
\usepackage{umoline}
\renewcommand{\Re}{\mathrm{Re}}

\definecolor{gray}{rgb}{.6,.6,.6}

\usepackage[normalem]{ulem}
\newcommand{\affiliationIKER}{\affiliation{$^{1}$IKERBASQUE, Basque Foundation for Science, 48013 Bilbao, Spain}}
\newcommand{\affiliationDIPC}{\affiliation{$^{2}$Donostia International Physics Center, 20018 San Sebasti\'an, Spain}}

\begin{document}
\title{Ferroelectric nano-traps for polar molecules}
\author{Omjyoti Dutta$^{1}$ and G. Giedke$^{1,2}$}
\affiliationDIPC
\affiliationIKER

\date{\today}

\begin{abstract}
  We propose and analyze an electrostatic-optical nano-scale trap for cold
  diatomic polar molecules. The main ingredient of our proposal is an
  square-array of ferroelectric nano-rods {with alternating polarization}. We show that, in
contrast to electrostatic traps using the linear 
  Stark effect, a quadratic Stark potential supports long-lived trapped
  states. The molecules are kept at a fixed height from the nano-rods by a
  standing-wave optical dipole trap. For the molecules and materials
  considered, we find that nano-traps with trap frequency up to 1MHz,
  ground-state width $\sim20$nm with lattice periodicity of $\sim 200$nm. 
  Analyzing the loss mechanisms due to non-adiabaticity,
  surface-induced radiative transitions, and laser-induced transitions, we
  show the existence of trapped states with life-time $\sim 1$s,
  competitive with current traps created via optical mechanisms. As an
  application we extend our discussion to an 1D array of nano-traps to
  simulate of a long-range spin Hamiltonian in our structure.
\end{abstract}

\maketitle
\section{Introduction}
Ultracold atoms and molecules trapped in optical potentials constitute a
versatile toolbox for simulating a plethora of Hamiltonians \cite{maciej}. The
energy scale for such trapping is set by the optical wavelength and laser
strength. To go beyond this energy scale, there is a recent surge in
investigations of trapping atoms in sub-wavelength lattices. Most of these
studies concentrate on hybrid atom-dielectric systems and the use of vacuum
forces to achieve lattice constants on the order of $\sim 50$nm \cite{chang,
  tudela}. Naturally, a pertinent question in this regard is to extend such
trapping schemes for polar molecules, which due to
  their rich internal structure and potentially strong interactions have
  raised considerable interest as a basis for quantum computation and
  simulation \cite{DeMille02,YKC06,BD+Zoller07}. Additionally, 
the presence of a permanent dipolar moment in such molecules is
responsible for a plethora of exotic  physics with applications in
quantum engineering \cite{dutta,pup,ye} and precision measurements
\cite{elecdipole}. The use of a nanoscale trap can be beneficial for these
 applications due to increased energy scale. For a  comparison, in typical 
optical lattices ($\lambda \sim 1.06\mu$m) with
a microwave coupled RbCs molecule, the maximum nearest-neighbour interaction energy 
is on the order of $\sim 0.5$ kHz ($30$nK). Since the dipolar interaction falls off as a cubic power of  
distance and a three-fold decrease in lattice constant will result in a 27fold increase of 
interaction strength. This energy scale can potentially give rise to $\sim 10^4$ gate 
operations, within a typical molecule lifetime of $\sim 1$ second, for
quantum {information processing} applications. Another {advantageous consequence} of nano-scale
confinement {originating from} reduced tunneling and 
overlap between neighbouring sites, a possible mechanism for
suppression of molecular complex formation \cite{bohnc}. 
We {propose a setup} to create a
sub-wavelength trap for {cold} rovibrational ground-state polar molecules which 
are {prepared, e.g.,} in optical lattices \cite{chotia,nagerl}. Ferroelectric materials can provide a natural 
{basis} for such traps. Nanoscale ferroelectricity is a source of intense research due to its potential application as non-volatile memories, sensors etc.\cite{fridkin}. It has been found that monodomain ferroelectricity survives for nano-rods with radius down to $\sim 20$nm \cite{Morozovska, Geneste, handbook}. Taking advantage of state-of-art lithography and nanotechnology \cite{nanoarray,hal,nanoconfine,nanopillar,nanotube,nanoreview, pztnano} 
techniques, it is possible to create a periodic array of ferroelectric nano-rods. Moreover, using cantilever tips
or external loads, polarization of each nano-cell can also be controlled 
externally \cite{cantilever,mechanical}. {This enables, in principle, the design of potential
landscapes to trap polar molecules.}

In this paper, we propose an electrostatic-optical trap for rigid-rotor $^{1}\Sigma$ diatomic
polar molecules. The sub-wavelength trap is provided by the electric field created by a periodic arrangement of
ferroelectric nano-rods. An optical potential is used to prevent the molecules
from moving away from the nanostructure. We show that it is possible to obtain a
nanometer-sized trap for molecules in high-field seeking states with a lifetime on the
order of seconds. Such life time is achieved by the virtue of two
ingredients: (i) the existence of a
trapped state with negligible non-adabatic (Majorana) loss and
(ii) the suppression of additional losses due to hyperfine mixing by
applying a strong magnetic field on the order of few Tesla for $1$D
trapping. 

At this point, we like to stress
that trapping molecules using electrostatic force has a long history
\cite{meijer}. In most of these studies, the molecules are trapped using
linear Stark shift. Such traps suffer from Majorana losses near the trap
center owing to their kinetic origin \cite{bohn}. Therefore, reducing the trap size to the nanometer (nm) regime in general 
will make the molecules extremely short-lived (lifetime on the order of few microseconds or less). In our proposal,
the trapping potential originates as a second order perturbative
effect (quadratic Stark shift). By analyzing the quantum motion, we show the
presence of motional states in which the non-adiabatic coupling is
weakened considerably by destructive interference. Additionally, we discuss molecule loss due to both vacuum photons and
thermal phonons in presence of a substrate. We finish the article by a
proposal to simulate a long-range $XX$ spin Hamiltonian. 

The paper is arranged as follows: We present the hyperfine and rotational structure of
a $^1\Sigma$ molecule in Sec.~\ref{sec:mol}, in particular, we consider $^{87}$Rb$^{133}$Cs
(RbCs) as a paradigmatic example. In Sec.~\ref{intro} we present our
ferroelectric structure corresponding to a $0$D arrangement of
nano-rods. Sec.~\ref{0D} contains a detailed analysis of the $0$D electric field
and its trapped states. In Sec.\ref{lassec} we close the trap in
  $Z$-direction by adding a suitable laser field. The main loss mechanisms
  (non-adiabatic and hyperfine-induced losses and losses due to nanostructure and laser field) are
discussed and shown to be small in Secs.~\ref{nonad} and
\ref{sec:surfloss}. The final Secs.~\ref{1dtrap} and \ref{qsim} consider the
extension to a 1D system and propose a quantum simulator
using our traps.

\section{Molecular Hamiltonian}\label{sec:mol}
Here we consider a rigid-rotor $^1\Sigma$ diatomic molecule
amenable to laser cooling. The molecular Hamiltonian in
the electronic and vibrational ground state is
given by \cite{brown},
\begin{eqnarray}\label{hyperfine}
H_{\rm mol}&=&\hbar B_e \bm{N}^2 + \sum_{i}c_i \bm{I}_i\cdot\bm{N} + c_4 \bm{I}_1\cdot \bm{I}_2 + H_{\rm Q} + H_{\rm mag}, \nonumber\\
H_{\rm Q} &=& \frac{\sum_i(eqQ)_i\left[ 3(\bm{I}_i\cdot \bm{N})^2 + \frac{3}{2}(\bm{I}_i\cdot \bm{N}) - \bm{N}^2\bm{I}^2_i\right]}{2\mathcal{I}_i(2\mathcal{I}_i-1)(2\mathcal{N}-1)(2\mathcal{N}+3)} , \nonumber\\
H_{\rm mag} &=& - g_r\mu_{\rm N} \bm{N}\cdot \vec{B} - \mu_{\rm N}\sum_i
g_i(1-\sigma_i) \bm{I}_i\cdot \vec{B}, 
\end{eqnarray}
where the first term gives the rotational spectrum with 
$\bm{N}$ is the total angular momentum operator and  $B_e$ is the rotational
constant. The second term takes into account interaction of the rotational angular
momentum with the nuclear spins and the third term denotes the 
direct interaction between the nuclear spins. As we are interested in diatomic molecules, the total nuclear spin operators of the two atoms are denoted by $\bm{I}_{1,2}$. The Hamiltonian $H_{\rm Q}$ denotes quadrupole interaction with coupling constants $(eqQ)_{1,2}$. The last term in the Hamiltonian denotes the Zeeman term in presence of a magnetic field $\vec{B}$ where $g_r$ is the rotational g-factor of the molecule and $g_{1,2}$ are the nuclear g-factors with typically $g_{1,2} \gg g_r$. In the present paper, we apply a magnetic field along the $Z$-direction, $\vec{B}=B_0\hat{Z}$. A full description of the internal molecular state is then denoted by $\ket{\mathcal{N},\mathcal{M_N},\mathcal{I}_1,\mathcal{M}_{\mathcal{I}_1},\mathcal{I}_2,\mathcal{M}_{\mathcal{I}_1}}$, where,
$\bm{N}^2 \ket{\mathcal{N},\mathcal{M_N}} =  \mathcal{N}(\mathcal{N}+1)\ket{\mathcal{N},\mathcal{M_N}}, \bm{N}_Z \ket{\mathcal{N},\mathcal{M_N}} = \mathcal{M_N} \ket{\mathcal{N},\mathcal{M_N}}, \mathcal{N} \in [0,1,2,\cdots], -\mathcal{N} \geq \mathcal{M_N} \geq \mathcal{N}$ and same for the nuclear spin operators.

As a first step, we diagonalize the Hamiltonian in Eq.~\eqref{hyperfine}
for the $^{87}$Rb$^{133}$Cs molecule ($\mathcal{I}_1=\mathcal{I}_{\rm Cs}=7/2,
\mathcal{I}_2=\mathcal{I}_{\rm Rb}=3/2$). The hyperfine constants are taken
from Ref. \cite{Hutsonhype}. We are specially interested in the rotational
levels $\mathcal{N}=1, \mathcal{M_N}=\pm1$. This
justification for choosing such state will be clarified in
Sec.~\ref{0D}. Moreover, for the present purpose  we ignore coupling to the $
\mathcal{N}=1, \mathcal{M_N}=0$ state which can be detuned by
application of laser fields (see Sec.~\ref{lassec}). For concreteness, we
chose the magnetic field {$B=B_0$} such that the {two lowest energy states are almost}  degenerate.These states are:
\begin{align}\label{magstates0}
\ket{\alpha_0} &= \ket{1,1,7/2,3/2}, \nonumber\\
\ket{\beta_0} &\approx  (1-\delta^2/2) \ket{1,-1,7/2,3/2} - \delta \ket{1,1,7/2,-1/2},
\end{align}
where $\delta \ll 1$, and $\ket{\alpha_0}$ in the maximally polarized state
which has no quadrupolar coupling. On the other hand, the $\ket{1,-1,7/2,3/2}$
state is coupled to $\ket{1,1,7/2,-1/2}$ via $H_{\rm Q}$. Without the magnetic
field such coupling will lead to an equal superposition of these
states. Moreover, the $H_{\rm mol}$-eigenvalues for the states
$\ket{\alpha_0},\ket{\beta_0}$ are $\approx2\hbar B_e -
E_0-\Delta_{\rm hf}/2$, and $\approx2\hbar B_e - E_0+\Delta_{\rm
  hf}/2$, where $E_0$ is the energy and $\Delta_{\rm hf}$ is
the detuning due to hyperfine and Zeeman Hamiltonian of Eq.~\eqref{hyperfine} .

For future use (see Sec.~\ref{0D}), we will also consider the
energy eigenstates {in which}
$\ket{1,{\pm}1,7/2,-1/2}$ has the majority contribution: 
\begin{eqnarray}\label{magstates1}
\ket{\alpha_1} &\approx & 1-\delta^2/2-\delta^2_1/2\ket{1,1,7/2,-1/2} + 
\delta_1\ket{1,1,5/2,1/2} \nonumber\\
&+& \delta\ket{1,-1,7/2,3/2}, \nonumber\\
\ket{\beta_1} &\approx & (1-\delta^2_1{/2})\ket{1,-1,7/2,-1/2} + \delta_1\ket{1,-1,5/2,1/2}, \nonumber\\
&&
\end{eqnarray}
where $\delta_1 \ll 1$ and $\ket{\alpha_1},
\ket{\beta_1}$ are $H_{\mathrm{mol}}$-eigenstates with eigenenergies
$2\hbar B_e - E_1 -\Delta_{\rm hf}/2\approx 2\hbar
B_e - E_1+\Delta_{\rm hf}/2$, where $E_1$ is the energy contribution
from the hyperfine and Zeeman Hamiltonian. In Sec.~\ref{0D} we will
see that these states are coupled to $\ket{\alpha_0}, \ket{\beta_0}$
by the electric field of the ferroelectric rods. For future use, we
introduce a shorthand notation 
for the nuclear spin quantum numbers by 
the collective symbol: $\mathcal{I}_{\rm col}\equiv \{\mathcal{I}_{\rm
  Cs},\mathcal{M}_{\mathcal{I}_{\rm Cs}},\mathcal{I}_{\rm
  Rb},\mathcal{M}_{\mathcal{I}_{\rm Rb}}\}$ and as a result a general state
is written as
\begin{equation} \label{ketdef}
\ket{\mathcal{N},\mathcal{M_N},\mathcal{I}_{\rm col}} \equiv \ket{\mathcal{N},\mathcal{M_N},\mathcal{I}_1,\mathcal{M}_{\mathcal{I}_1},\mathcal{I}_2,\mathcal{M}_{\mathcal{I}_1}}.
\end{equation}

\section{Polar-molecules near $0$D ferroelectric nano-structures}
\label{intro}
\begin{figure}
    \includegraphics[width=0.5\textwidth]{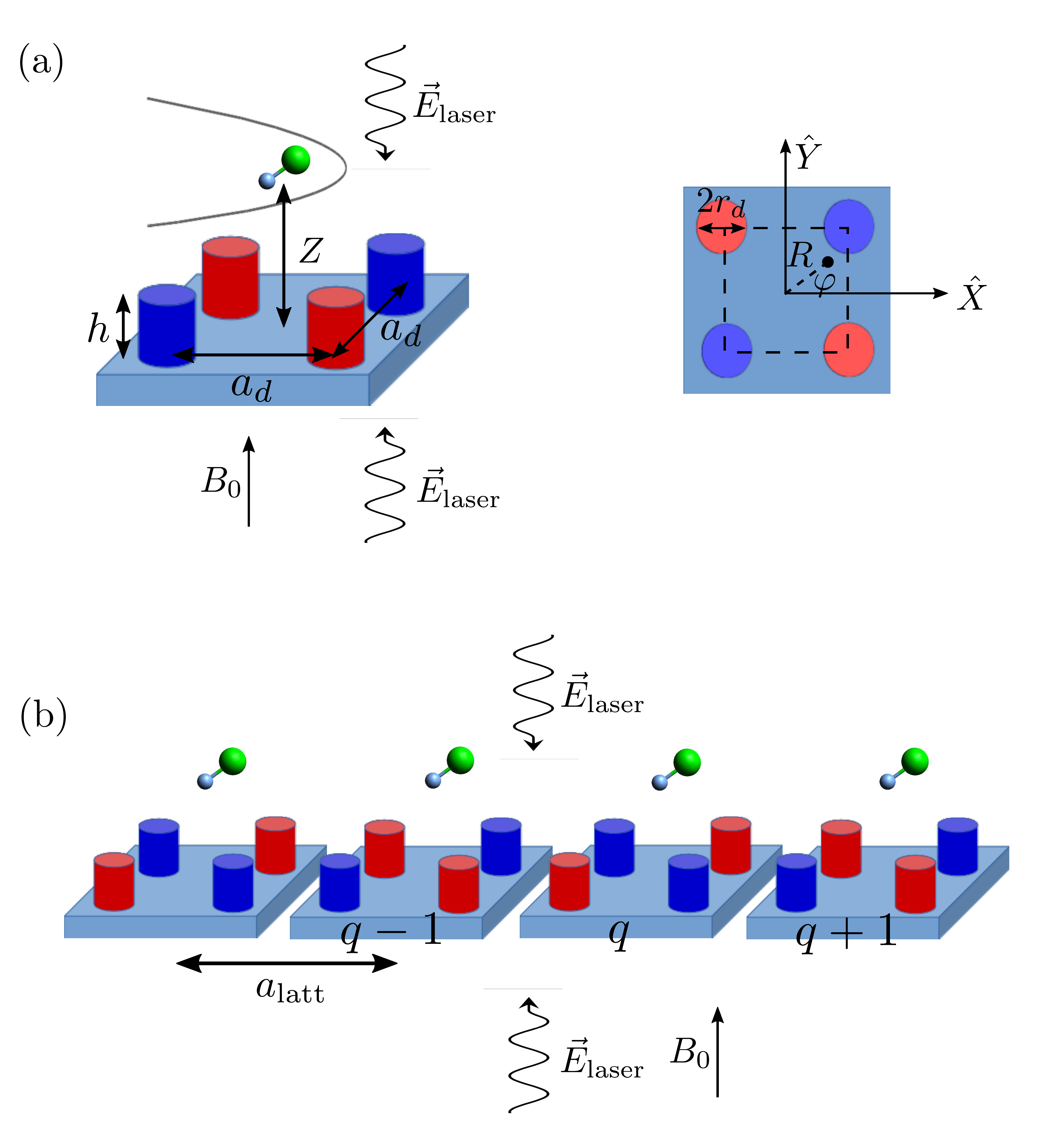}
    \caption{ (a) Left panel: Side view of our primitive cell is shown. Each
      nano-rod is made of ferroelectric material. The red and blue
      colour of the rods denote the $\pm \hat{Z}$ direction of the
      ferroelectric polarization, respectively. We place a dipolar
      molecule  at a height $Z$ above the top of the nano-rods. We illuminate the system by counter-propagating optical laser
      fields, which provide a trap for the molecule along the
      $Z$-direction. Moreover, we apply a static magnetic field
      $B_0\hat{Z}$ to minimize hyperfine loss. Right panel:
      Top view of our primitive cell. We define the local polar
      co-ordinate system. (b) We arrange the primitive cells
      periodically along the $X$-axis. Notice the $\pi/2$ rotation of
      the ferroelectric arrangement between neighbouring cells. The
      periodicity of the cells is denoted by $a_{\rm latt}$.}   
        \label{figure1}
\end{figure}

Our system consists of a symmetric arrangement of 
polarized ferroelectric 
nano-rods. The elementary structure consists of four cylindrical nano-rods centered
around each corner of a square as shown in Fig.~\ref{figure1}(a). Each
nano-rod has a radius $r_{\rm d}$ and height $h\geq r_{\rm d}$. Within a cell, neighbouring nano-rods are separated by a distance $a_d$. The
unit vectors $\hat{X}$ and $\hat{Y}$ along the plane of the cell are
shown in {the second panel of} Fig.~\ref{figure1}(a). We measure $Z$ from the top surface of
the nano-rods {and denote the centers of the four rods by
  $(m_x,m_y)a_d/2\equiv\vec{F}_{\mathbf{m}}, m_i=\pm1$,
  respectively}. We consider an 
anti-ferroelectric arrangement where 
polarization of each rod
is given by {$P_{\mathbf{m}}=(-1)^{(m_x+m_y)/2}P$}. From
now on, we scale all distances by the nano-rod radius $r_{\rm d}$
unless otherwise explicitly specified. In this paper such nano-rod arrangement is referred to as $0$D structure.

We study polar molecules near such ferroelectric
nano-structures. In general, our model Hamiltonian is given by $H_{\rm
  sys}=H_{\rm kin}+H_{\rm mol}+H_{\rm mf}$, where $H_{\rm kin}$ is the kinetic
energy of the center of mass of the molecule, $H_{\rm mol}$ is given in Eq.~(\ref{hyperfine}) and $H_{\rm mf}$ denotes the
interaction between a single molecule and a nano-structured cylinder.

%---------------------------
%----------------------------

\subsection{Molecule-Ferroelectric interaction}

First, we set the notation for the molecular position as $\rho =
(\vec{R},Z)$, where the
transverse vector $\vec{R}\equiv(X,Y)$. The molecule-ferroelectric
interaction is given by the dipole Hamiltonian and expressed as,
\begin{equation}\label{molferro}
H_{\rm mf} = -\alpha_{\rm mf} \left[E_Z[\vec{\rho}] \bm{T^1_0} + E_{-}[\vec{\rho}]
\frac{\bm{T^1_1}}{\sqrt{2}} + E_{+}[\vec{\rho}] \frac{\bm{T^1_{-1}}}{\sqrt{2}} \right],
\end{equation}
where $\bm{T}^1_0=\cos\theta_{\rm in}$, $\bm{T}^1_{\pm 1}= \sin\theta_{\rm in}
 e^{\pm i\phi_{\rm in}}/\sqrt{2}$ are the spherical tensors of rank 1
denoting the internal co-ordinates of the molecule and only act on the rotational states $\ket{\mathcal{N},\mathcal{M_N}}$. The molecular axis in the laboratory frame is defined by the angles to the $Z$-axis and its projection to he $X$-axis, $\theta_{\rm
 in},\phi_{\rm in}$. Here, $E_Z[\vec{\rho}]$ is the $Z$-component of the electric field and $E_{\pm}[\vec{\rho}]=(E_X[\vec{\rho}]\pm iE_Y[\vec{\rho}])$ are its azimuthal components. We should note that electric fields are dimensionless in our units. We define the effective molecule-ferroelectric coupling strength,
$\alpha_{\rm mf} = \frac{\mu P}{4\pi\epsilon_0}$, where $\mu$ is the
permanent dipole moment of the molecule. For RbCs molecules, the
dipole moment is given by $\mu=1.22$ Debye \cite{Hutson}. 

After carrying out integration over the height of the nano-rods, the field
strength at $\vec{\rho}$ due to the ferroelectric island at
position $(m_xa_d,m_ya_d)$ reads  
\begin{align}\label{efield-indiv}
E_Z[\bm{m};\vec{\rho}] &=-\int d\vec{r} \left(ZE[Z]-(Z+h)E[Z+h]\right), \nonumber\\
E_{-}[\bm{m};\vec{\rho}] &= -  e^{-i\tilde{\phi}_{\bm{m}}} \int d\vec{r} \left(E[Z+h]-E[Z]\right) \\
& \times 
\left({R}_{\bm{m}}-r e^{-i(\phi-{\phi}_{\bm{m}})}\right),  \nonumber\\
E[Z] & =\frac{1}{{\left(Z^2+|\vec{R}_{\mathbf{m}}-\vec{r}|^2\right)^{3/2}}}\nonumber
\end{align}
where $\vec{R}_{\bm{m}}=\vec{R}-\vec{F}_{\bm{m}}$, $\bm{m} \equiv (m_x,m_y)$ and
$\tan{\phi}_{\bm{m}}=(Y-m_ya_d)/(X-m_xa_d)$ and we have defined a
local polar coordinate around each nano-rod axis $\vec{r}\equiv (r,\phi)$. The integral is defined as $\int
d\vec{r} =\int^{1}_{0}rdr\int^{2\pi}_0 d\phi$. The total
field components are given as the sum of the contributions of all
nano-rods by  $E_Z[\vec{\rho}], E_{-}[\vec{\rho}]$:
\begin{align}
  \label{efield}
  E_{\eta}[\vec{\rho}]&=\sum_{\bm{m}}(-1)^{\frac{m_x+m_y}{2}}E_{\eta}[\bm{m};\vec{\rho}] \,\,\,(\eta=Z, -).
\end{align}
 For the azimuthal electric fields$E_{+}[\vec{\rho}]=E^{*}_{-}[\vec{\rho}]$. From
 the expression in Eq.~\eqref{efield-indiv}, it is clear that each
 ferroelectric nano-rod can be effectively substituted by
 {opposite} surface charges at the top and bottom of the nano-rods.

To treat the effect of ferroelectric nano-rods {on the molecular
  motion} and identify the
  trap provided by the proposed arrangement, we operate in a regime with $|E_{\pm}[\vec{\rho}]|, |E_{Z}[\vec{\rho}]| \ll B_e$, which allows us to carry out a second order perturbation analysis for
each $\mathcal{N}$ manifold. Note that selection rules imply that
$\bra{{\cal N}}H_{\rm mf}\ket{{\cal N}'}\propto\delta_{{\cal N}'{\cal
    N}\pm1}$. Moreover, we neglect the detuning effect of hyperfine
splitting as $c_{1,2},c_4, (eqQ)_{1,2}, \mu_{\rm N}B_0 \ll B_e$ and
suppress the  hyperfine structures of the molecular state in the
notation. From  the Hamiltonians in Eqs.~(\ref{hyperfine},\ref{molferro}), we derive an effective potential of the form
\begin{eqnarray}\label{effpot1}
V^{\mathcal{N}} &=& \frac{\alpha^2_{\rm mf}}{2\hbar B_e}\left(E_{+}[\vec{\rho}]E_{-}[\vec{\rho}]V^{\mathcal{N}}_{\perp}+E^2_{-}[\vec{\rho}]V^{\mathcal{N}}_{+} + E^2_{+}[\vec{\rho}]V^{\mathcal{N}}_{-} \right.\nonumber\\ 
&+& \left. E^2_{Z}[\vec{\rho}]V^{\mathcal{N}}_{Z} + E_{Z}[\vec{\rho}]E_{-}[\vec{\rho}]V^{\mathcal{N}}_{Z+} + E_{Z}[\vec{\rho}]E_{+}[\vec{\rho}]V^{\mathcal{N}}_{Z-}\right), \nonumber\\
&&
\end{eqnarray}
where $V^{\mathcal{N}}_{-}=[V^{\mathcal{N}}_{+}]^\dagger, V^{\mathcal{N}}_{Z-}=[V^{\mathcal{N}}_{Z+}]^\dagger$ and
\begin{eqnarray}
V^{\mathcal{N}}_{\perp}&=&\sum_{\eta=\pm 1}\sum_{\mathcal{N}',\mathcal{M_{N'}}}\frac{\bm{T^1_\eta}\ket{\mathcal{N}',\mathcal{M_{N'}}}\bra{\mathcal{N}',\mathcal{M_{N'}}}\bm{T^1_{-\eta}}}{E_\mathcal{N}-E_\mathcal{N'}},\nonumber\\
V^{\mathcal{N}}_{+}&=&\sum_{\mathcal{N}',\mathcal{M_{N'}}}\frac{\bm{T^1_{1}}\ket{\mathcal{N}',\mathcal{M_{N'}}}\bra{\mathcal{N}',\mathcal{M_{N'}}}\bm{T^1_{1}}}{E_\mathcal{N}-E_\mathcal{N'}}, \nonumber\\
V^{\mathcal{N}}_{Z}&=&\sum_{\mathcal{N}',\mathcal{M_{N'}}}\frac{\bm{T^1_0}\ket{\mathcal{N}',\mathcal{M_{N'}}}\bra{\mathcal{N}',\mathcal{M_{N'}}}\bm{T^1_{0}}}{E_\mathcal{N}-E_\mathcal{N'}},\nonumber\\
V^{\mathcal{N}}_{Z+}&=&\sum_{\substack{\eta,\eta'=0,1 \\ \eta \neq \eta'}}\sum_{\mathcal{N}',\mathcal{M_{N'}}}\frac{\bm{T^1_\eta}\ket{\mathcal{N}',\mathcal{M_{N'}}}\bra{\mathcal{N}',\mathcal{M_{N'}}}\bm{T^1_{\eta'}}}{E_\mathcal{N}-E_\mathcal{N'}},\nonumber\\
&&
\end{eqnarray}
While deriving Eq.~\eqref{effpot1}, we have made the assumption that
the molecular center of mass motion is adiabatic, i.e., it is slow
compared to the rotational splitting. In App.~\ref{app1}, we discuss the non-adiabatic corrections to this approximation. 

\section{$0$D nano-traps for molecules}
\label{0D}
First we investigate the simplest ferroelectric geometry consisting of
only four nano-rods, i.e., {the 0D structure introduced above}, cf.\ Fig.~\ref{figure1}. As we shall see,
this configuration can laterally confine molecules in a suitable
internal state to the center of the arrangement, hence we call it a
$0$D nano-trap.
To obtain this result, we consider the form of the electric field at
the center of the square. We can gain useful insights from 
  symmetry arguments: the field sources change orientation under rotation
  (around the $Z$-axis) by $\pi/2$ as well as under reflections either
  at $Y=0$ or at $X=0$. 
This constrains the Fourier series of the azimuthal field component
(written using in-plane polar coordinates $R,\varphi$, collectively referred to as $\vec{R}$)
$E_-(\varphi,R,Z)=\sum_l c_l(R,Z)e^{il\varphi}$ (cf. App.~\ref{app:symmetry}):  only the coefficients
$c_l$ with $l=4j+1$ may be non-zero and they are imaginary and odd functions of
$R$. Specifically, we have that $c_{-1}[R,Z]=c_{3}[R,Z]=0$. 

\begin{figure*}
    \includegraphics[width=.8\textwidth]{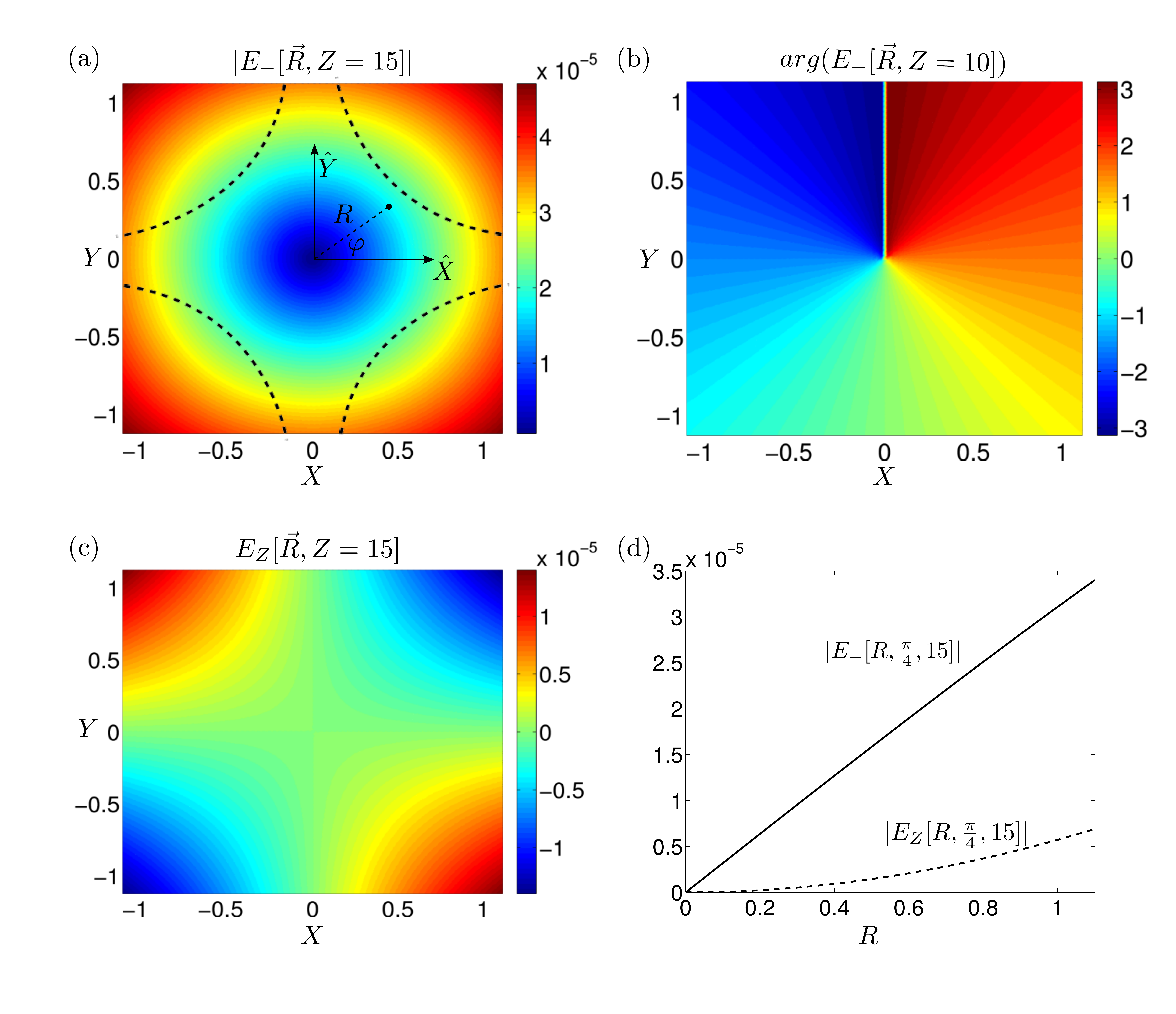}
    \caption{ Electric field distribution of the $0$D nano-rods
      arrangement as in Fig.~\ref{figure1} with $N_f=0$ and $a_d=2.25$. (a) The magnitude of the azimuthal
      field distribution $E_{-}[\vec{\rho}]$ in the $XY$-plane is
      shown for $Z=15$. At the center of the square, electric field
      vanishes due to the reflection- and $\pi/2$ rotational-
      symmetries. Moreover, we define a local polar co-ordinate from
      the center of each square: $(R,\varphi)$. For $R\lesssim 1$, the
      $E_[\vec{\rho}]$ field magnitude grows linearly with $R$. The
      black dashed lines denote boundaries of the nano-rods along
      $XY$-plane. (b) The argument of $E_{-}[\vec{\rho}]$ is shown. We clearly see that in terms of the local polar co-ordinate, $E_{-}[\vec{\rho}]\approx ie^{i\varphi}$. (c) The field distribution of $E_Z$ is plotted. The $Z$-field has maximum magnitude along the two diagonals and has a angular distribution $\propto \sin 2\varphi$.
    (d) $|E_{-}[\vec{\rho}], E_{Z}[\vec{\rho}]|$ are plotted as a
    function of distance along the diagonal from the center.} 
        \label{figure2}
\end{figure*}

\begin{figure*}
    \includegraphics[width=1\textwidth]{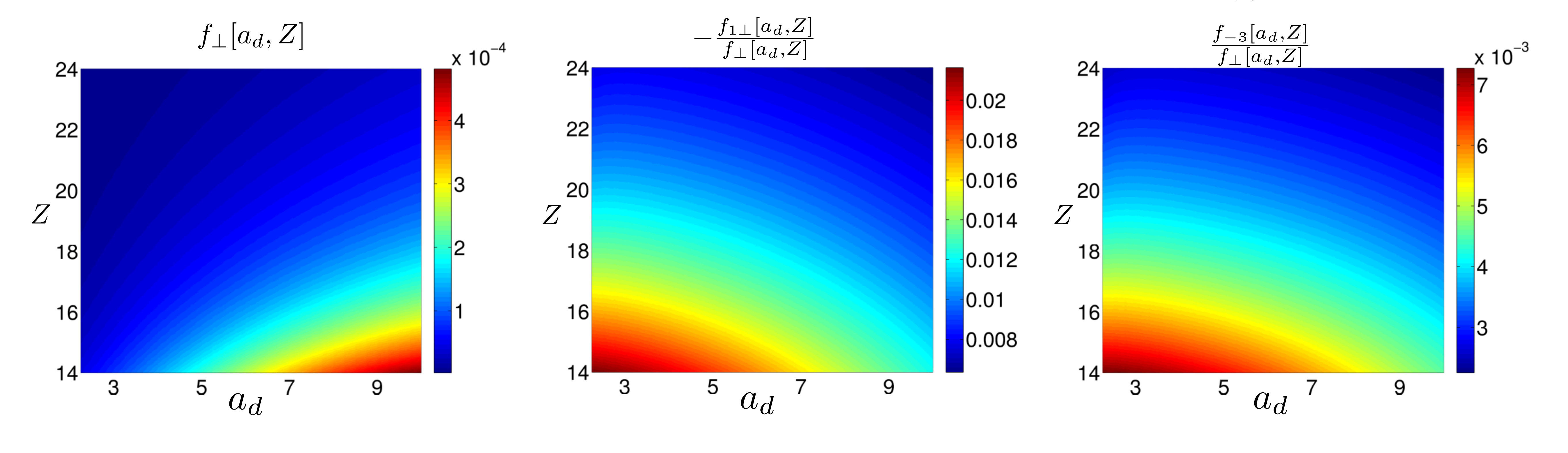}
    \caption{We plot the fitting functions in Eq.~(\ref{series0d}) as a function of $a_d$ and $Z$. Left panel: $f_{\perp}[a_d,Z]$, Middle panel: $-f_{1\perp}[a_d,Z]/f_{\perp}[a_d,Z]$, and Right panel: $f_{-3}[a_d,Z]/f_{\perp}[a_d,Z]$.} 
        \label{figure2a}
\end{figure*}

Moreover, note that as the electric field is non-singular, we
have in general that $c_n[R,Z] \propto R^n$ as $R\rightarrow 0$. 
Thus $c_{1}[R,Z]\propto R$ for small $R$. For the
$Z$-components, we can similarly show that it contains only even
powers of $R$ and by Gauss' law $\nabla\cdot\vec{E}=0$ we have
$E_Z[\varphi,R,Z] \propto \mathcal{O}[R^2]$ for small $R$.
Such electric fields resemble a traditional quadrupolar field configuration.
For illustration purposes, we plot the field distributions in
Fig.~\ref{figure2}. In Fig.~\ref{figure2}(a,b) amplitude and argument
of the azimuthal field is shown. For $R < a_d/2$, we indeed notice
that $E_{-}[\vec{\rho}]\propto ie^{i\varphi}$. The linear dependence of $E_{-}[\vec{\rho}]$ with $R$ is clearly seen in Fig.~\ref{figure2}(d) as is the angular dependence of $E_Z[\vec{\rho}]$ in Eq.~\eqref{series0d} in Fig.~\ref{figure2}(c). Moreover, $E_Z[\vec{\rho}]$ depends quadratically on $R$ to the leading order as can be verified from Fig.~\ref{figure2}(d). 
 %---------------------------
%----------------------------
Thus we find for small $R$ for the field given in Eq.~\eqref{efield} that
\begin{eqnarray}\label{series0d}
E_{-}[\vec{\rho}] &\approx& iR
(f_{\perp}[a_d,Z]+f_{1\perp}[a_d,Z] R^2)\exp[i \varphi]\nonumber\\
&+& if_{-3}[a_d,Z]\exp[-3i\varphi]R^3 + \mathcal{O}[R^5], \nonumber\\
E_{Z}[\vec{\rho}] &\approx& f_{z}[a_d,Z]R^2\sin [2\varphi] + \mathcal{O}[R^4], 
\end{eqnarray}
where Gauss' law leads to the form of $E_Z[\vec{\rho}]$ with
$\partial_Zf_z[a_d,Z]-3(2f_{1\perp}[a_d,Z]+f_{-3}[a_d,Z])=0$. We use
$f_{\perp}[a_d,Z], f_{1\perp}[a_d,Z], f_{-3}[a_d,Z]$ as fitting
functions which we obtain numerically and which are shown in Fig.~\ref{figure2a}. It is clear that the largest parameter is
 %---------------------------
%----------------------------
$f_{\perp}[a_d,Z]$ corresponding to the linear dependence of the
azimuthal electric field with $R$ as
$f_{1\perp}[a_d,Z]/f_{\perp}[a_d,Z] \sim 10^{-2}$, and
$f_{-3}[a_d,Z]/f_{\perp}[a_d,Z] \sim 5.0\cdot 10^{-3}$. Moreover, we
see that by increasing the lattice constant one increases $f_{\perp}[a_d,Z]$ and simultaneously one decreases the relative strength of the fitting functions associated with $R^3$ scaling.  Whereas, by increasing $Z$, we see a decrease in each fitting function due to the scaling of the dipolar interaction. 

As the electric fields vanish as $R\rightarrow0$, we can write the
effect of the ferroelectric nano-rods within perturbation theory as
presented in Eq.~\eqref{effpot1}. We consider the first excited
rotational level $\mathcal{N}=1$ manifold with $\mathcal{M}_1= \pm
1$. The  $\mathcal{M_N}=0$ state is excluded {from further
  consideration} due to its large shift in
energy from the $\mathcal{M_N}= \pm 1$ as will be shown
later. Moreover, for $R < a_d/2$, the $E_{\pm}[\vec{\rho}]$ field components
are dominant compared to $E_Z[\vec{\rho}]$ which is then neglected at
first. Subsequently we only keep the $V^{1}_{\perp},V^{1}_{\pm}$ from
Eq.~\eqref{effpot1}.  Using  the state description of
Eq.~\eqref{ketdef}, we find that the effective potential is
\begin{eqnarray}\label{trapham}
V^{1}_{\rm rad}&=& \frac{f^2_{\perp}[a_d,Z] \alpha^2_{\rm mf}}{40\hbar B_e} R^2 \sum_{\substack{\mathcal{M}_1=\pm 1 \\ \mathcal{I}_{\rm col}}}\left( \ket{1,\mathcal{M}_1,\mathcal{I}_{\rm col}}\bra{1,\mathcal{M}_1,\mathcal{I}_{\rm col}} \right. \nonumber\\
 &-& \left.{3} e^{2i\mathcal{M}_1\varphi}\ket{1,\mathcal{M}_1,\mathcal{I}_{\rm col}}\bra{1,-\mathcal{M}_1,\mathcal{I}_{\rm col}}\right),
\end{eqnarray}
where we have used the expression for $E_{\pm}[\vec{\rho}]$ from Eq.~\eqref{series0d} retaining the leading term $\propto R$.  By inspecting
Eq.~\eqref{trapham}, it is clear that for each $\mathcal{I}_{\rm
    col}$ a trapped state can be formed by
the superposition: 
\begin{equation}\label{trapsup}
\ket{-,\varphi}\equiv \sum_{\mathcal{M}_1=\pm 1} \ket{1,\mathcal{M}_1,\mathcal{I}_{\rm col}}e^{i\mathcal{M}_1\varphi}/\sqrt{2} 
\end{equation} 

We are now ready to consider the joint effect of the internal Hamiltonian $H_{\rm mol}$ in Eq.~\eqref{hyperfine} and the effective potential in Eq.~\eqref{trapham}. Specifically, we will transform to the diagonal basis of $H_{\rm mol}$, and consider the states $\ket{\alpha_{0,1}}, \ket{\beta_{0,1}}$ in Eqs.~\eqref{magstates0}, \eqref{magstates1}.

Exploiting the approximate cylindrical symmetry of our problem, we
write down explicitly the position dependence of individual internal
states as, $\braket{\vec{R}\mid \alpha_j}=\sum_{\ell} a^j_{\ell}[R]
e^{-i\ell\varphi}/\sqrt{2\pi}$ and $\braket{\vec{R}\mid
  \beta_j}=\sum_{\ell} b^j_{\ell}[R] e^{-i\ell\varphi}/\sqrt{2\pi}$,
where $j=0,1$ denotes the molecular internal state and
$\ell=0,\pm1,\pm2,\cdots$ represents the center-of-mass angular
momentum around the laboratory $Z$-axis. Additionally, for the time
being, we neglect motion of the particle along the $Z$-axis
(see Sec.\ref{lassec}) and as a result the
$Z$-dependence is implicit in the coefficients. We use the transformed coefficients: $t^j_{\ell}[R]=(a^j_{\ell}[R]-b^j_{\ell-2}[R])/\sqrt{2}, u_\ell[R]=(a^j_{\ell}[R]+b^j_{\ell-2}[R])/\sqrt{2}$. Moreover to simplify the notation, we introduce the array: $\bm{t}^j_\ell[R] = [t^j_{\ell}[R] \hspace{0.1cm} u^j_{\ell}[R]]^{T}$. In terms of the transformed coefficients, the
Schr\"{o}dinger equation in closed form reads
\begin{eqnarray}\label{eigeneq21}
\begin{bmatrix}
H_{00}&  H_{01} \\
H_{10} & H_{11}
\end{bmatrix}
\begin{bmatrix}
\bm{t}^0_\ell[R]\\
\bm{t}^1_{\ell-2}[R]
\end{bmatrix}&=&
\epsilon \begin{bmatrix}
\bm{t}^0_\ell[R]\\
\bm{t}^1_{\ell-2}[R]
\end{bmatrix},
\end{eqnarray}
where the matrices are given by
\begin{widetext}
\begin{eqnarray}\label{eigeneq22}
H_{00} &=& \begin{bmatrix}
- K\left(\partial^2_R + \frac{\partial_R}{R} - \frac{(\ell-1)^2+1}{R^2}\right) + \frac{(f_{\perp}[a_d,Z]  \alpha_{\rm mf})^2R^2}{10\hbar B_e} - E_0&  2K\frac{\ell-1}{R^2}-\frac{\Delta_{\rm hf}}{2}\\
2K\frac{\ell-1}{R^2}-\frac{\Delta_{\rm hf}}{2} &  -K \left(\partial^2_R + \frac{\partial_R}{R} - \frac{(\ell-1)^2+1}{R^2}\right) - \frac{(f_{\perp}[a_d,Z] \alpha_{\rm mf})^2R^2}{20\hbar B_e}- E_0
\end{bmatrix}, \nonumber\\
H_{11} &=& \begin{bmatrix}
-K \left(\partial^2_R + \frac{\partial_R}{R} - \frac{(\ell-3)^2+1}{R^2}\right) + \frac{(f_{\perp}[a_d,Z]  \alpha_{\rm mf})^2R^2}{10\hbar B_e} - E_1 &  2T\frac{\ell-3}{R^2}-\frac{\Delta_{\rm hf}}{2}\\
2K\frac{\ell-3}{R^2}-\frac{\Delta_{\rm hf}}{2} &  -K \left(\partial^2_R + \frac{\partial_R}{R} - \frac{(\ell-1)^2+1}{R^2}\right) - \frac{(f_{\perp}[a_d,Z] \alpha_{\rm mf})R^2}{20\hbar B_e}-E_1
\end{bmatrix},\nonumber\\
H_{01} &=&-\frac{3\delta(f_{\perp}[a_d,Z]  \alpha_{\rm mf})^2R^2}{40\hbar B_e} \begin{bmatrix}
0 & 1 \\
1 &  0
\end{bmatrix},
\end{eqnarray}
and $H_{10}=H_{01}$, where we have introduced the unit for kinetic
energy $K=\frac{\hbar^2}{2m_{\rm mol}r^2_d}$ and neglected terms with
strength $\propto \delta^2$. To solve
Eqs.~(\ref{eigeneq21},\ref{eigeneq22}), as a first approximation we
neglect the off-diagonal elements $H_{01}, H_{10}$ and $\Delta_{\rm
  hf}$, since $\delta \ll 1$. Later on in Sec.~\ref{nonad} we consider
the consequence of $H_{01}, H_{10}$, and $\Delta_{\rm hf}\not=0$ on the trap lifetime. Within this approximation, the Schr\"{o}dinger equation becomes 
\begin{eqnarray}\label{trapdiag}
\begin{bmatrix}
-K \left(\partial^2_R + \frac{\partial_R}{R} - \frac{(\ell-1)^2+1}{R^2}\right) + \frac{(f_{\perp}[a_d,Z]  \alpha_{\rm mf})^2R^2}{10\hbar B_e} - E_j&  2K\frac{\ell-1}{R^2}\\
2K\frac{\ell-1}{R^2} &  -K \left(\partial^2_R + \frac{\partial_R}{R} - \frac{(\ell-1)^2+1}{R^2}\right) - \frac{(f_{\perp}[a_d,Z] \alpha_{\rm mf})^2R^2}{20\hbar B_e}- E_j
\end{bmatrix} &=& \epsilon \begin{bmatrix}
{t}^j_\ell[R]\\
{u}^j_{\ell}[R]
\end{bmatrix}, \nonumber\\
&&
\end{eqnarray}
\end{widetext}
Note that the nano-rod-dependent potential term (that depends on
$f_\perp$) describes a harmonic trap for the upper part of
Eq.~(\ref{trapdiag}), while it comes with the opposite sign in the
lower block. Consequently, we divide Eq.~\eqref{trapdiag} in
two parts, the diagonal part where $t^j_\ell[R]$ denotes bound
(trapped) states due to the $R^2$ potential and $u^j_\ell[R]$ denotes untrapped states
with positive energy ($\epsilon >0$). A coupling between the two given
by the off-diagonal term. We notice in Eq.~\eqref{trapdiag} that apart
from the diagonal energy shift $E_j$, the equation of motion is
independent of the hyperfine internal index $j$. For $\ell=1$ the
off-diagonal part vanishes. Neglecting it at first for all $\ell$, the
differential equation for $t^j_\ell[R]\equiv t_\ell[R]$ can be transformed to associated
Laguerre equations with known solutions. The eigenvectors and
eigenenergies are given by
\begin{align}\label{oscen}
\ket{t^j;\ell;N;\mathcal{N}=1} &= e^{i(\ell-1)\varphi} \ket{t,\ell N}\ket{-,\varphi}, \nonumber\\ 
\braket{\vec{\tilde{R}}\mid t,\ell N}&= \frac{\sqrt{2}\left(\tilde{R}\right)^{\ell_{\rm eff}}\exp\left[-\tilde{R}^2/2\right]\mathcal{L}^{\ell_{\rm eff}}_N\left[\tilde{R}^2\right]}{(\Gamma[N+\ell_{\rm eff}+1]/N!)^{1/2}}, \nonumber\\
\epsilon_{\ell, N}[a_d,Z]&= (2N+\ell_{\rm eff}+1)\hbar\omega[a_d,Z],
\end{align}
where 
\begin{align}
  \label{eq:4}
  \tilde{R}&=R/\sigma[a_d,Z],\\
 \ell_{\rm eff}&=\sqrt{(\ell-1)^2+1} 
\end{align}
and $\Gamma[\cdots]$ is the Gamma
function. The quantum number of the radial motion
of the trapped molecules is represented by non-negative integers $N$,
with $N=0$ representing the lowest energy state. The $Z$-dependent
oscillator width ($\sigma[a_d,Z]$) and frequency $\omega[a_d,Z]$ have the form:
\begin{eqnarray}\label{osc}
\sigma^2[a_d,Z]&\equiv&\left(\sigma[Z]\right)^2=\frac{\left(10 K \hbar
  B_e\right)^{1/2}}{f_{\perp}[a_d,Z] \alpha_{\rm mf}}, \nonumber\\ 
\hbar\omega[a_d,Z] &=& 
|\alpha_{\rm mf}|f_{\perp}[a_d,Z] \left(\frac{2K}{5\hbar B_e}\right)^{1/2}.
\end{eqnarray}
From Eq.~\eqref{oscen} it is clear that $\ell=1$ is lowest in energy
and is non-degenerate. All other levels are two-fold degenerate between the pair of states, $\ket{t^j;-\ell+2;N;\mathcal{N}=1}, \ket{t^j;\ell;N;\mathcal{N}=1}, \ell>1$. Looking into the radial distribution of the trapped states in Eq.~\eqref{oscen}, we see that the wavefunction vanishes at $R=0$. As a result the effect of the cross-term in Eq.~\eqref{trapdiag} will be small. Moreover, for  $N=0$, the excitation energy from the lowest energy state is given by $\delta\epsilon=
\epsilon_{0, 0}[a_d,Z]-\epsilon_{1, 0}[a_d,Z]= \epsilon_{2, 0}[a_d,Z]-\epsilon_{1, 0}[a_d,Z] =(\sqrt{2}-1)\hbar\omega[a_d,Z]$.
\begin{table}[h!]
  \centering
   \caption{Ferroelectric parameters for $0$D and $1$D trap}
   \label{tab:table1}      
   \begin{tabular}{c | c c c c c c}
   \toprule
   & $P$(C$\cdot$m$^{-2}$) & $r_d$(nm) & $a_{\rm latt}(nm)$ & $h_d$(nm) & $\frac{\alpha_{\rm mf}}{\hbar B_e}$ & $\frac{K}{\hbar B_e}$ \\
   \midrule
   $0$D & $10^{-1}$ & $60$ & $\infty$ &$180$ & $5.8 \cdot 10^{3}$ & $1.2 \cdot 10^{-5}$ \\
   \midrule
   $1$D & $2.5\cdot10^{-1}$ & $45$ & $200$ & $135$ & $2.9 \cdot 10^{4}$ & $2.2 \cdot 10^{-5}$ \\
   \bottomrule
   \end{tabular}  
  \vspace{0.25cm}  
  \caption{Molecular parameters for RbCs}   
  \label{tab:table2}
  \begin{tabular}{c c c}
     \toprule
      $\mu$(Debye) & $m_{\rm mol}$($10^{-26}$kg) & $B_e$(GHz)\\
     \midrule
     $1.22$ & $37$ & $0.5$\\
     \bottomrule
  \end{tabular}
  \caption{Parameters representing molecular states in Eqs.~(\ref{magstates0}, \ref{magstates1}) and the corresponding magnetic field and energy scales}   
    \label{tab:table3}
    \begin{tabular}{c | c c c c c}
       \toprule
       & $B_0$(T) & $\delta$ & $\delta_1$ & $\frac{E_1-E_0}{\hbar B_e}$& $\frac{\Delta_{\rm hf}}{1\hbar B_e}$\\
       \midrule
       $0$D & $0.1$ & $0.054$ & $0.058$ & $5.5\cdot 10^{-3}$ & $0.0$ \\
       \bottomrule
       $1$D & $2.0$ & $0.003$ & $0.003$ & $5.5\cdot 10^{-1}$ & $3.7\cdot 10^{-4}$ \\
    \end{tabular}
  
\end{table}
Next, we discuss the properties of the trap. It is clear that as the
molecule is held closer to the surface, the trapping frequency
increases as noted in Fig.~\ref{figure3}(a). For such states to
be trapped along $Z$, one needs an external force to keep the molecule
near the nano-structure; this matter will be addressed in the next
section. Additionally, by increasing the lattice constant $a_d$ upto a certain value, one can
also increase the trap frequency. 

To qualitatively characterize the number of trapped states present, first we define the effective potential without the quadratic approximation as
$V_{\rm trap}[\vec{\rho}]=\frac{|E_{-}[\vec{\rho}]|^2
 \alpha^2_{\rm mf}}{10\hbar B_e}$. As seen from Fig.~\ref{figure2}(a), the trap height is minimal along $\varphi=0$ or $\pi/2$, and consequently the trap depth is defined as
\begin{equation}
  \label{eq:trapdepth}
  V_{\rm depth}[a_d,Z]=V_{\rm trap}[R=a_d/2,\varphi=0,Z]. 
\end{equation}
Then the quantity $N_{\rm max}=V_{\rm
  depth}[a_d,Z]/2\hbar\omega[a_d,Z]$ gives an estimate for the number
of trapped states; it is plotted in Fig.~\ref{figure3}(b). We see that
as the lattice constant decreases, the trap becomes shallow. For a
fixed lattice constant, as expected, bringing the molecule closer to
the surface results in an increased number of trapped
states. Approximately, the trap ceases to exist as $Z$ increases and
$\hbar\omega[a_d,Z] \approx V_{\rm depth}[a_d,Z]$ is reached. For the
parameters given in Tables~\ref{tab:table1} and \ref{tab:table2}, and
using Eq.~\eqref{osc}, we find that {for $a_d=2.25$ and at $Z=16$} the trapping frequency is
$\omega[2.25,16]\approx 0.5$MHz, cf. Fig.~\ref{figure3}(a). 

\section{Trapping the molecules along $Z$ direction}
\label{lassec}
From the trapping energy Eq.~\eqref{oscen}, it can be noticed that as
the molecule in the (laterally) trapped state moves closer to the
surface, the trap energy increases. Hence, the molecules will be pushed
away from the surface. To prevent such an escape, we locally trap the
molecules along the $Z$-direction by employing a standing-wave optical
laser field far red-detuned from the excited electronic states. Via the AC
Stark effect this attracts the molecule to the high-intensity region of the
beam. However, the presence of such a field can lead to loss of
molecules as components of the laser field polarized in the
 $XY$-plane will strongly mix the trapped and untrapped internal
states \cite{jin} described above. To prevent such polarization loss, we
propose to use a
light beam with a focused waist along the $Y$-direction and a longitudinal
component along the $Z$-direction. Moreover, the longitudinal component
needs to be much stronger than the transverse component in the trapping
region. Furthermore, we keep in mind that we want to generalize our trapping
geometry from $0$D to $1$D. Which means that the property of the beam should
be approximately unaltered by translation along either $X$- or $Y$-axis. To
realize such a beam, we pass a Hermite-Gaussian wave 
through a cylindrical lens to focus at $Z=Z_0$ along the $Y$-axis. As
a zeroth order approximation \cite{dispnano}, 
($a_d/\lambda \ll 1$, where $\lambda$ is the laser wavelength) 
we neglect the effect of 
the nanostructure on the laser field at the molecular position so long as
$\lambda/a_d > n$, where $n$ is the refractive
index of the ferroelectric substrate. This condition is equivalent to
the physical situation that only the zeroth order diffraction mode
exists and the coupling to the guided modes of the periodic dielectric
system is minimal due to normal incidence. For the incident
field, we assume $\mathcal{E}^{\pm}_{\rm
  inc}=\sqrt{P_{in}}(0,2Y/w_0,0) \exp[-Y^2/w^2_0 \pm i kZ-i\Omega_{\rm
  las}t]$, where $P_{\rm in}$ denotes the laser power, $w_0$ the beam waist, 
and $k=2\pi/\lambda$. Such a mode can be created by passing a Gaussian
wave through a $\pi$ phase plate \cite{light1}. Along the
$X$-direction we have assumed a uniform field distribution. The
time-independent contribution of the light field after focusing
through a cylindrical lens (cylinder axis along $\hat{X}$) is given by
$\mathcal{E}^{\pm}_{\rm
  las}=(0,\mathcal{E}^{\pm}_Y,\mathcal{E}^{\pm}_Z)$ (apart from a
position-independent phase factor) \cite{light1,light2}, 
\begin{eqnarray}\label{laserfield}
\mathcal{E}^{\pm}_Y[Y,Z] &=& \sqrt{P_{\rm in}}\frac{2kf^2}{w_0}\int_{\psi_1}^{-\psi_1}  F[\psi]\cos^{3/2}[\psi]\sin[\psi] \nonumber\\
&\times& \exp\left[\pm ikY\sin[\psi]\pm ik(Z\mp Z_0)\cos[\psi]\right] d\psi,\nonumber\\
\mathcal{E}^{\pm}_Z[Y,Z] &=& \sqrt{P_{\rm in}} \frac{2kf^2}{w_0} \int_{\psi_1}^{-\psi_1}  F[\psi]\cos^{1/2}[\psi]\sin^2[\psi] \\
&\times& \exp\left[\pm ikY\sin[\psi] \pm ik(Z\mp Z_0)\cos[\psi]\right] d\psi,\nonumber
\end{eqnarray}
where 
$F[\psi]=\exp\left[-f^2\sin[\psi]^2/w^2_0\right]$ is the window function of
the lens and $f$ is its focal length. The numerical aperture of the lens is given by $\sin[\psi_1]=a_0/f$ where $a_0$ is the width of the pupil. The total laser field is given by adding two counter-propagating fields in Eq.~\eqref{laserfield} and defined as $\mathcal{E}_{\rm tot}[Y,Z]=(0,\mathcal{E}^Y_{\rm tot}[Y,Z],\mathcal{E}^Z_{\rm tot}[Y,Z])$. The total $Z$-field 
%---------------------------
\begin{figure}
    \includegraphics[width=0.5\textwidth]{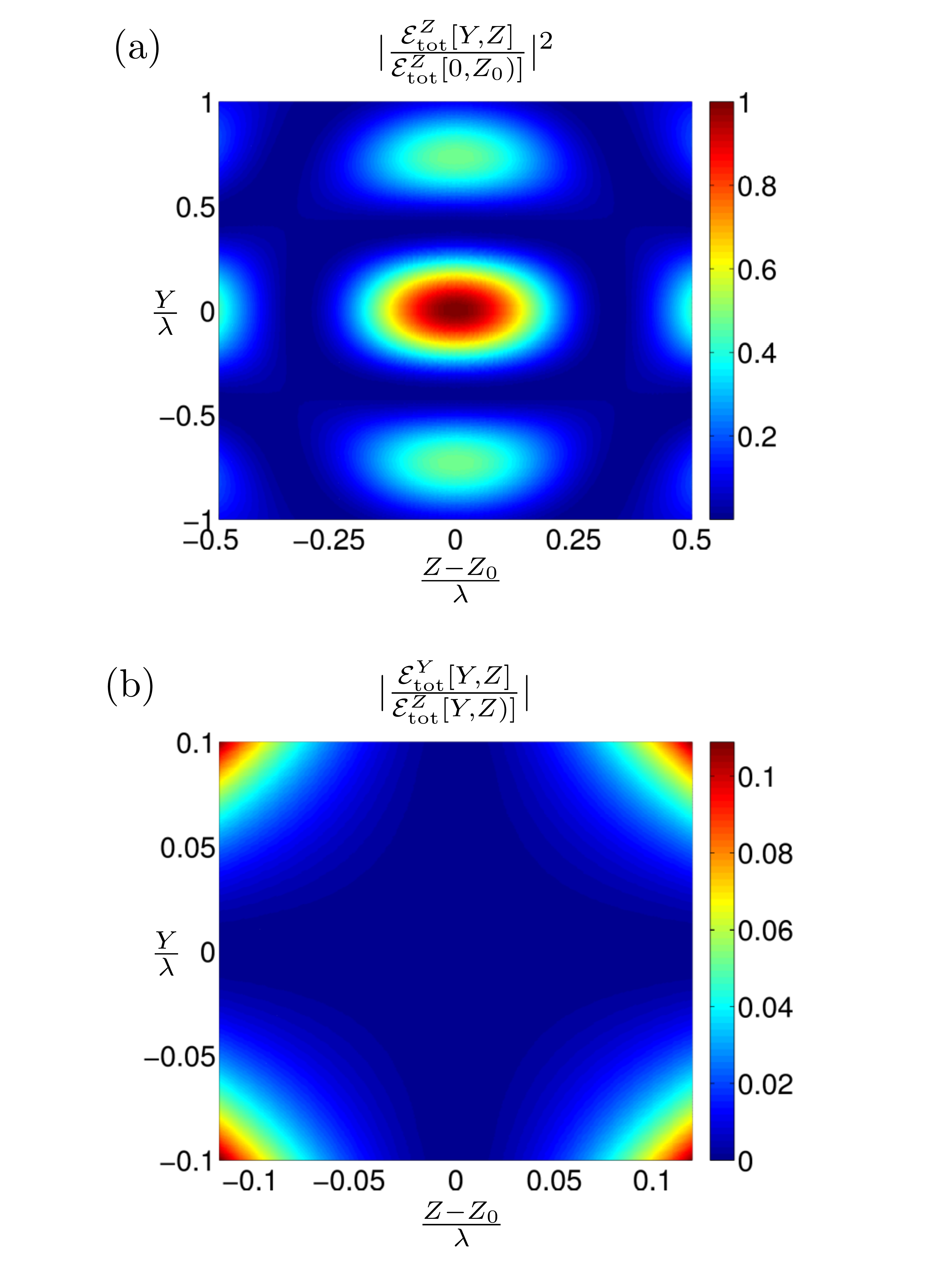}
    \caption{ (a) Plot of the longitudinal $Z$-component of the total electric field. (b) Ratio between the transverse $Y$-component and the longitudinal $Z$-component is shown. The limits on the axis are different in (a) and (b). The parameters are, $\psi_1=\pi/3, f/w_0=1.15$. } 
        \label{figure4}
\end{figure}
%----------------------------
shows a maximum at the focal line $Z=Z_0$, $Y=0$ as seen in Fig.~\ref{figure4}(a). From Eq.~\eqref{laserfield} it is clear that at $Y=0$, $\mathcal{E}^{\pm}_Y[Y,Z]=0$ as the integrand is odd under $\psi\rightarrow-\psi$. From Fig.~\ref{figure4}(b), we see that at the line $Z=Z_0,Y=0$ the $Y$ component of the total field vanishes. This is due to the out-of-phase oscillation of the $Y$ field as a function of $Z$. As a result, for $Z\approx Z_0,Y\rightarrow 0$, we are always in a region where the longitudinal $Z$-component is much stronger than the transverse component and for $Z\rightarrow Z_0,Y=0$, the laser intensity is approximately quadratic: $|\mathcal{E}^Z_{\rm tot}[0,Z]|^2 \approx I_0(1-\beta (Z-Z_0)^2/\lambda^2), \mathcal{E}^Y_{\rm tot}=0$, where $\beta \approx 4\pi^2 $ is a constant denoting curvature of the intensity profile near $Z=Z_0$ and $I_0$ is the total power of the lasers after focusing.

Next, we consider the effective laser-induced potential for $\mathcal{N}=1$
molecules. The laser-molecule interaction Hamiltonian projected onto the
$\mathcal{M}_{\mathcal{N}=1}\in\left\{ 0,\pm1
  \right\}$ subspace is given by
(apart from a constant shift in energy $=-\alpha_1 I_0$) \cite{kot,kot1,jin} by
\begin{align}\label{laspot}
\frac{V^1_{\rm light}}{I_0}&= -\alpha_0 \sum_{\mathcal{I}_{\rm col}}
\ket{1,0,\mathcal{I}_{\rm col}}\bra{1,0,\mathcal{I}_{\rm col}} + \nonumber\\ 
&+\alpha_1 V_{\rm Z} \sum_{\substack{\mathcal{M}_1=\pm 1 \\ \mathcal{I}_{\rm col}}} \ket{1,\mathcal{M}_1,\mathcal{I}_{\rm col}}\bra{1,\mathcal{M}_1,\mathcal{I}_{\rm col}}, 
\end{align}
where $V_{\rm Z}=\sin^2\left[\frac{2\pi(Z-Z_0)}{\lambda}\right]$, and
$\alpha_1$ is the polarizability of the $\ket{\pm 1}$ states for a
$Z$-polarized light field, which can be expressed as
$\alpha_1=(\alpha_{\parallel}+4\alpha_{\perp})/5$, where $\alpha_{\parallel},\alpha_{\perp}$ are anisotropic
polarizabilities of the molecule. Similarly,
$\alpha_0=2(\alpha_{\parallel}-\alpha_{\perp})/5$, which shows that the
$\mathcal{M}_1=0$ state will be detuned in energy from the $\mathcal{M}_1=\pm
1$ states. Moreover, we consider laser strengths such that the detuning
$\alpha_0I_0$ is much larger than the transverse trap frequency
$\hbar\omega[a_d,Z]$. For a laser strength of $I_0 \sim 0.1$MW$\cdot$cm$^{-2}$
and the trap frequency range in Fig.~\ref{figure3}(a),
$\alpha_0I_0/(\hbar\omega[a_d,Z]) \sim 20$ which will exponentially
suppress
the loss rate to the $\mathcal{M}_1=0$ state by an approximate factor
$e^{-\alpha_0I_0/(\hbar\omega[a_d,Z])}$. The exponential factor arises due to the overlap integral between the trapped and the continuum state with energy $\sim \alpha_0I_0$sim. Then, combining Eqs.~(\ref{oscen},\ref{laspot}), the total
effective potential along the $Z$-direction seen by the trapped state is given
by
\begin{eqnarray}\label{effp}
V^1_{\rm eff}[Z]&=&\epsilon_{\ell,N}[Z]+\alpha_1 I_0\sin^2\left[\frac{2\pi(Z-Z_0)}{\lambda}\right], \nonumber\\
  &\approx& V^1_{\rm eff}[Z_1] + \frac{d^2 V^1_{\rm eff}[Z]}{2dZ^2} {\big |}_{Z=Z_1} (Z-Z_1)^2,
\end{eqnarray}
where in the second line we have Taylor expanded the potential around the local minimum $Z_1$ (the subscript $1$ is to keep tab of the rotational level).
%---------------------------
%\begin{figure}
%    \includegraphics[width=0.5\textwidth]{figure5.pdf}
%    \caption{Effective potential of Eq.~\eqref{effp} for total laser intensities $I_0=1.0$(solid line)MW$\cdot$cm$^{-2}$ for $a_d=3.0$(dashed line), $3.5$ (dash-dotted line), $4.0$ (solid line).} 
%        \label{figure5}
%\end{figure}
%----------------------------
We find that for the region of $Z$ with minimized loss rate (discussed in the next section) the local minimum of Eq.~\eqref{effp} coincides with $Z_1 \approx Z_0$ for a laser with wavelength $\lambda=1090$nm with focal plane $Z_0 \sim 16$ and $I_0 \sim 0.1$MW$\cdot$cm$^{-2}$. The approximate ground state along the $Z$-direction is expressed as,
$$
\Phi_Z=\left(\frac{1}{\pi {\sigma_Z}^2}\right)^{1/4}\exp\left[-\frac{(Z-Z_0)^2}{2{\sigma_Z}^2}\right],
$$
where the wave-function width is given by
${\sigma_Z}^{2}=\left(\frac{d^2 V^1_{\rm eff}[Z]}{2dZ^2} {\big
|}_{Z=Z_0}\right)^{-1/2}$. For the parameters considered here, $\alpha_1 I_0 (r_d/\lambda)^2\gg \omega[a_d,Z]$ and as a result the trap frequency in Z direction is given by $\omega_Z=4\pi\sqrt{\alpha_1I_0K}(r_d/\lambda)$.
   
\textbf{\textit{Laser-induced loss --}} Due to the red-detuned nature of the
laser light, we trap the molecule at an intensity maximum. This can lead to
molecular loss due to the imaginary part of the polarizability \cite{kot}. For
a laser wavelength of $1\mu$m, the imaginary part is $10^{-7}-10^{-8}$ times
weaker than the real part. Thus for a laser intensity of $I_0 \sim
0.1$MW$\cdot$cm$^{-2}$ gives a lifetime in the order of $1-10$s. One way to
further increase the lifetime is by increasing the laser wavelength to around
$1.5\mu$m where one is off-resonant from all excited states and as a result
the imaginary part should decrease exponentially with respect to frequency
shift: As seen in Ref.~\cite{kot}, the real part of the polarizability remains the same, but the imaginary part decreases by an order of magnitude or more. As a result, one can increase the molecular lifetime to tens of seconds although the trapping along $Z$ direction becomes shallow.  

\section{Molecule loss rates due to non-adiabatic and hyperfine coupling}
\label{nonad}

Next, we discuss various couplings that can transfer trapped states to
untrapped states. The first approximation arises as a non-adiabatic effect to the Hamiltonian due to the position dependence of the perturbation in Eq. \eqref{effpot1}. As discussed in Appendix \ref{app1}, in the present case, such corrections are found to be negligible.

In the remaining part of this section, we describe important loss mechanisms.

\subsubsection{Loss due to intra-state coupling}
\begin{figure*}
    \includegraphics[width=1.0\textwidth]{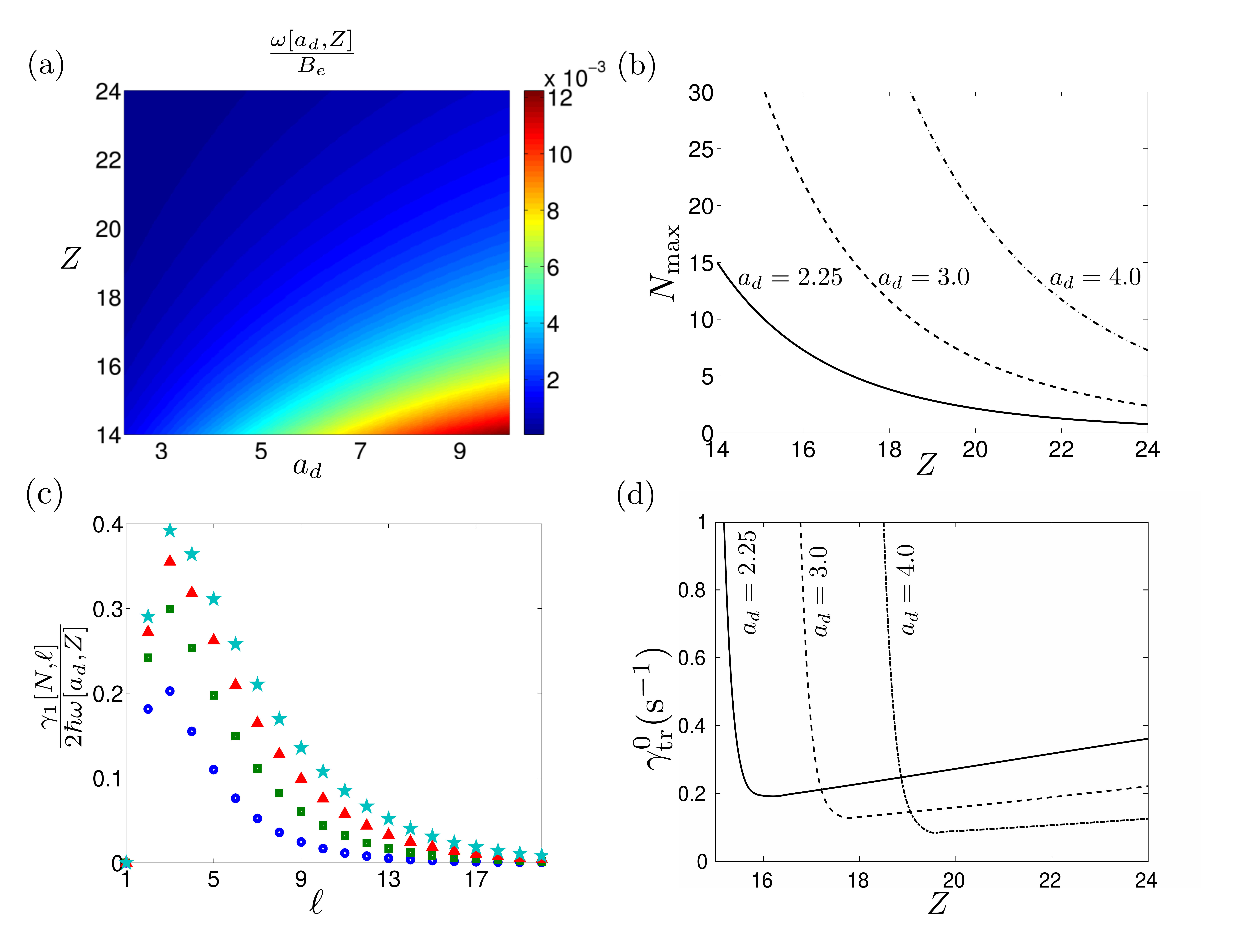}
    \caption{ (a) The trap frequency is shown as a function of $Z$ and
      $a_d$ for the parameters listed in Tables~\ref{tab:table1} and \ref{tab:table2}. (b) Relative trap depth $N_{\rm max}$ is shown as a function of $Z$ for $a_d=2.25$ (solid line), $3.0$ (dashed line), $4.0$ (dash-dotted line). (c) The decay rate $\gamma_1[\ell, N]$ from Eq.~\eqref{decay1} is plotted as a function of $\ell$ for $N=0(\circ), 1(\square), 2(\triangle), 3(\star)$. (d) Total non-adiabatic and hyperfine-induced loss rate $\gamma^0_{\rm tr}$ is plotted as a function of $Z$ for $a_d=2.25$ (solid line), $3.0$ (dashed line), $4.0$ (dash-dotted line). The  parameters are shown in Tables~\ref{tab:table1}, \ref{tab:table2}.} 
        \label{figure3}
\end{figure*}

The next correction arises due to the non-zero off-diagonal elements of
$H_{11}$ in Eq.~\eqref{trapdiag}. These elements couple the trapped state
$\ket{t^j;\ell;N;\mathcal{N}=1}$ to the continuum states $\ket{u^j;\ell;\mathcal{N}=1}$ with same angular momentum $\ell$ as described by the off-diagonal terms in Eq.~\eqref{trapdiag}. It is readily seen that a special situation arises for $\ell=1$ as the off-diagonal term
vanishes and there is no coupling to the untrapped state. This exact
decoupling of trapped and untrapped states no longer holds for
$\ell\not=1$. To qualitatively describe the effect of the untrapped states, we invoke Fermi's Golden Rule by
considering resonant coupling of the trapped state
$\ket{t^j;\ell;N;\mathcal{N}=1}$ to the continuum states with energy
$\epsilon_{\ell, N} > 0$ via the off-diagonal term in
Eq.~\eqref{trapdiag}. The solution for the untrapped state
then becomes: $\braket{\vec{\tilde{R}}\mid u,\ell M}=\sqrt{2}\mathcal{J}_{\ell_{\rm
    eff}}[\alpha_{\ell,M}\tilde{R}/\tilde{R}_0]/\left(\tilde{R}_0\left|\mathcal{J}_{\ell_{\rm
        eff}+1}[\alpha_{\ell,M}]\right|\right)$,
where $\mathcal{J}_\ell[]$ denotes the Bessel function of order
$\ell$. We use a cylindrical hard-wall boundary condition with radius
$\tilde{R}_0\rightarrow \infty$ and $\alpha_{\ell,M}$ is $M$th zero of the
Bessel function, $\mathcal{J}_{\ell_{\rm eff}}[x\alpha_{\ell,M}]=0$. The resonant energy
condition is given by 
$\alpha^2_{\ell,M_{\rm res}}/\tilde{R}^2_0=2(2N+\ell_{\rm eff}+1)+\tilde{R}^2/2$. 
The decay rates for the trapped state $\ket{t^j;\ell;N;\mathcal{N}=1}$ reads
\begin{eqnarray}\label{decay1}
\frac{{\gamma_1[\ell, N]}}{\hbar\omega[a_d,Z]} &=& 2\pi\mathcal{D}[\alpha_{\ell, M_{\rm res}}] \left|\braket{t,\ell N \mid \tilde{R}^{-2}\mid u, \ell M_{\rm res}}\right|^2,
\end{eqnarray}
where the density of states in dimensionless units is $\mathcal{D}[\alpha_{\ell,M}]={\tilde{R}_0}^2/(2\pi \alpha_{\ell,M})$. The use
of Fermi's Golden Rule remains valid as long as the decay rates are
lower than the minimum energy gap, $|\epsilon_{\ell,
  N+1}-\epsilon_{\ell, N}|=2\hbar \omega[a_d,Z]>\gamma_1[\ell,N]$, which is fulfilled for
all $\ell$. We plot the decay rate in Fig.~\ref{figure3}(c) for various $\ell, N$. 
%---------------------------
%----------------------------
Due to the symmetry around $\ell=1$,  
$\gamma[-\ell, N]=\gamma[\ell+2, N]$ for $\ell\geq 0$. We find that
the decay rate is maximal for $\ell=5,-3$ and then decreases for
larger $\ell$. Note that, as the effective potentials are always of finite height,  as a result the decay rates are valid as long as $\epsilon_{\ell, N} \lesssim V_{\rm deptth}$. 
We point out that trapped states with this kind of long lifetime
($\ket{t;\ell=1;N;\mathcal{N}=1}$) do not exist for traps using the linear
Stark shift (e.g., for asymmetric-top molecules). As a result, for
those traps, one
needs a larger trap size (which decreases $K$) to suppress molecule loss. The reason behind this is that there exists no angular momentum channel for which the coupling to untrapped states vanishes.

\subsubsection{Loss due to hyperfine structure induced coupling}\label{subs:hfloss}
Hyperfine induced coupling has two contribution. The first one induced intra-state transition due to the presence of detuning $\Delta_{\rm hf}$ in $H_{00}$ and $H_{01}$ in Eq.~\eqref{eigeneq22}. From Table \ref{tab:table3}) and Fig.~\ref{figure3}(a) we find that the trap frequency is much larger than the detuning ($\Delta_{\rm hf}/\omega[a_d,Z] \approx 0$) and as a result we neglect its effect. The next correction arises due to the matrix elements of
$H_{01}, H_{10}$ in Eq.~\eqref{eigeneq22} which are of order
  $\delta\ll 1$ (cf. Table \ref{tab:table3}) and induce coupling between  different internal states
denoted by $j$. As the 
off-diagonal elements of $H_{01}, H_{10}$ are non-zero, these elements can
couple a trapped state to continuum states belonging to a
different hyperfine structure. On the hand, there is a detunning due to the presence of magnetic field with magnitude $|E_0-E_1|$. As a result, for a sufficiently strong magnetic field, the transition is suppressed for $|E_0-E_1|/\omega[a_d,Z]\gg 1$. Such a suppression occurs as
the trapped $\ket{t^0, 10}$ state is confined deep inside the
classically forbidden region of the continuum states $\ket{u^1,-1}$.

Though for magnetic fields available in a laboratory, one can not reach a regime of complete suppression. As a result, we calculate the transition rate from the trapped state (for details see Appendix \ref{app:hyperfine})
which we denote by $\gamma^j_{\rm hf}$. There definition is given in Eqs.~(\ref{decaystate0}, \ref{decaystate1}). The most important thing to notice by dimensional analysis of $H_{01}, H_{10}$ that $\gamma^j_{\rm hf} \propto \delta^2 \omega[a_d,Z]$. 

\subsubsection{Loss rate due to the $R^3$-dependence of electric field} 

Additional loss channels are also present due to the second term $\propto R^3\exp[-3i\varphi]$ in Eq.~\eqref{series0d} which modifies the $E_\pm$ electric field components. As a result, the effective potential in Eq.~\eqref{trapham} will be modified with an additional term $\propto R^4$. As shown in Appendix ~\ref{app:r3loss}, the correction leads to coupling within the same hyperfine manifold $j$ between states belonging to different $\ell$ quantum number. The modified decay rate due to coupling of $\ell=1, N=0$ motional state to other lossy trapped state is given by (for details see Appendix \ref{app:r3loss})
\begin{align}\label{decaytrapr4}
\gamma_t &= \left(\frac{f_{-3\perp}[a_d,Z]K }{f_{\perp}[a_d,Z]\hbar\omega[a_d,Z]}\right)^2\sum_{\ell'=-3,5 }\sum_{N'} \nonumber\\
&\times \left|\frac{V_{1,0;\ell',N'}}{-2N'+1-\sqrt{(\ell'-1)^2+1}}\right|^2{\gamma_1[\ell', N']}
\end{align}
We like to point out that the use of perturbation theory may become
invalid for calculating the decay for states with higher $\ell$ due to
the presence of nearby degenerate states. In that case, one can get
the decay rates by concentrating on the degenerate
subspace. If that is not the case, we find that
the sum in Eq.~\eqref{decaytrapr4} approximately converges for
$|N'|<2$. As a result, for a consistent decay
rate from $N=0$ state, one needs to have a trap with $N_{\rm max} \sim
2$. Otherwise, one also need to consider the continuum states due to the finite hight of the trapped potential.

The next source of loss originates from coupling of the $\ell=1$ state to  continuum states with different $\ell$ quantum number as shown in Appendix \ref{app:r3loss}. The loss rate for the $N=0,\ell=1$ state consequently is given by,
\begin{eqnarray}\label{decaycontr4}
\gamma_{\rm c} &=& 2\pi \left(\frac{3f_{-3}[a_d,Z]}{4f_{\perp}[a_d,Z]\hbar\omega[a_d,Z]}\right)^2\frac{K^2\mathcal{D}[\alpha_{\ell, M_{\rm res}}]}{\hbar\omega[a_d,Z]}\nonumber\\
&\times& \sum_{\ell=-3,5}\left|\braket{t,\ell N \mid \tilde{R}^{4}\mid u, \ell M_{\rm res}}\right|^2,
\end{eqnarray}
with the resonant energy condition
$\alpha^2_{\ell,M_{\rm res}}/\tilde{R}^2_0=2(\ell_{\rm
  eff}+1)+\tilde{R}^2/2$.
To find the life-time, we assume the parameters as in Tables
\ref{tab:table1}, \ref{tab:table2} and the unperturbed loss rate from Eq.~\eqref{decay1}. 

\subsubsection{Total trapped molecule loss rate}
To find the total loss rate, we notice from Eqs.~\eqref{decaycontr4},
\eqref{decaytrapr4}, and subsection \ref{subs:hfloss} that the hyperfine-induced rate of level $j$ scales
as $\gamma^j_{\rm hf} \propto \omega[a_d,Z]$. On the other hand, from
Eqs.~(\ref{decaytrapr4}, \ref{decaycontr4}) and Eq.~(\ref{decay1}) we find that higher order
correction to the electric field gives rise to loss rates
$\gamma_{t,c} \propto \omega^{-1}[a_d,Z]$. As a result, once we fix the
ferroelectric polarization and the nano-rod dimensions, by changing
the position of the molecule along $Z$, one can find an optimum
solution. To this end, we define the total molecule loss rate by adding Eqs.~\ref{decaytrapr4}, \ref{decaycontr4}) and hyperfine-induced loss rate,
\begin{equation}
  \label{eq:10}
  \gamma^j_{\rm tr} = \gamma^j_{\rm hf} + \gamma_t+\gamma_c, 
\end{equation}
which is pictorially shown in Fig.~\ref{figure3}(d). From
Figs.~\ref{figure3}(b, d), we find that for $a_d=2.25$ ($a_d=135nm$ for parameters in Table:~\ref{tab:table1}) and with $N_{\max} \approx 10$, the loss rate is minimized around a distance $Z \approx 16.0$
($\approx 0.96\mu$m) with decay rate $\gamma_{\rm tr}^0\approx0.2$Hz (lifetime of $\sim 5$s). By increasing the distance between the nano-rods,
one sees that the lifetime is increased upto $\sim 10$ seconds for $a_d=4.0$. Another way to increase the lifetime is by increasing the magnetic field which will increase $|E_1-E_0|$ between the hyperfine states $j=0,1$ and decreasing the hyperfine loss rates in subsection \ref{subs:hfloss}. In Appendix \ref{app: hyperfinepic} we show the effect of increased magnetic field with increased lifetime of $\sim 20$s for $Z=14$. 

We find that the remaining non-adiabatic loss channels due to the
$E_{\rm Z}$ field in Eq.~\eqref{effpot1} give a much longer lifetime
and as a result they are discussed in Apps.~\ref{app:ezfield1} and \ref{app:ezfield2}.

\section{Loss due to surface proximity} \label{sec:surfloss}
In the present section concerns loss of molecule from its rotational state due to thermalization in presence of the substrate surface. Such losses exists irrespective of the presence or absence of trapping potential. Such loss rates comes from two primary sources: i) photon fluctuations of the vacuum-substrate interface and ii) phonon fluctuations in the surface of the substrate.
    
\textbf{\textit{Radiative loss --}} A source of loss of molecules is
the coupling of rotational levels to the blackbody radiation modified
by nano-rods and surface of the 2D substrate. 
The coupling frequency then corresponds to a rotational transition which for $^1\Sigma$ molecules generally lies in the GHz region and subsequently we neglect the hyperfine splitting. The coupling wavelength corresponding to $\lambda_{\rm rot} \sim 10^{-1}$m. The height of
the nano-rods is negligible compared to the coupling wavelength,
$h/\lambda_{\rm rot} \ll1$. As a result from the viewpoint of
effective medium theory, the Fresnel reflection coefficients only get
modified by a negligible amount, $\propto h/\lambda_{\rm
  rot}$ \cite{effm}, and we can neglect the effect of the periodic
nano-rods. Moreover, the lifetime of the molecule is dominated by the
surface and, as a result, we neglect the free-space
contribution. Assuming that the substrate- and free-space photons are in
equilibrium with temperature $k_bT \gg \hbar B_e$, and following
\cite{heat}, the rotational heating rate for a molecule from the
$\ket{\mathcal{N}=1,\mathcal{M_N}=\pm 1}$ state is given by
\begin{eqnarray}
  \label{eq:1}
  \gamma_h &=& \gamma_0+\frac{\mu^2}{8\pi\epsilon_0 Z^3_{\rm mol-sub}} \frac{k_BT}{2\hbar B_e} \frac{Qf}{2B_e}\frac{\Re[\epsilon_{s}]}{(\Re[\epsilon_{s}]+1)^2} \nonumber\\
  &\times& \left(\braket{D^{-}_{00}}^2+\frac{\braket{D^{+}_{22}}^2}{2^2}+\frac{\braket{D^{-}_{20}}^2}{2^2}+\frac{\braket{D^{Z}_{21}}^2}{2}\right),
\end{eqnarray}
where $\gamma_0$  is the free-space heating rate, $Z_{\rm mol-sub}$ is the distance of the molecule from
the substrate, $\epsilon_{s}$ the dielectric constant of the
substrate, and the $D_{ij}$ are dipole matrix elements
\begin{align*}
\braket{D^{-}_{00}}&=\bra{\mathcal{N}=0,\mathcal{M}_\mathcal{N}=0}\bm{T^1_{-1}}\ket{\mathcal{N}=1,\mathcal{M}_\mathcal{N}=1},\\ 
\braket{D^{+}_{22}}&=\bra{\mathcal{N}=2,\mathcal{M}_\mathcal{N}=2
}\bm{T^1_{1}}\ket{\mathcal{N}=1,\mathcal{M}_\mathcal{N}=1},\\ 
\braket{D^{-}_{20}}&=\bra{\mathcal{N}=2,\mathcal{M}_\mathcal{N}=0
}\bm{T^1_{-1}}\ket{\mathcal{N}=1,\mathcal{M}_\mathcal{N}=1},\\ 
\braket{D^{Z}_{21}}&=\bra{\mathcal{N}=2,\mathcal{M}_\mathcal{N}=1 }\bm{T^1_{0}}\ket{\mathcal{N}=1,\mathcal{M}_\mathcal{N}=1}.
\end{align*}
Hence, the total heating rate is given given by the sum of the
free-space and substrate-induced heating rates \cite{heat}. The lifetime of a RbCs molecule at $4$K is on the order of $\sim 10^8$s \cite{heat} as a result we can practically neglect the free-space heating rate compared to the substrate-induced rate. To estimate the latter, we use fused quartz as an example whose dielectric properties are given by \cite{diesilica},
$\Re[\epsilon_{s}] \approx 3.83$ and $Qf\approx10^5$GHz. The
molecular heating rate then becomes $\gamma_{\rm h} \approx 0.02$s$^{-1}$
at a distance of $1\mu$m (equivalent to $Z\approx 16$ in the unit of
nano-rod radius for the parameters in Table \ref{tab:table1}) from the substrate for liquid Helium temperature of $4$K. The role of Casimir forces in such distance is negligible and is discussed
qualitatively in Appendix \ref{app:casimir}. 

One possible way to extend the lifetime can be achieved by a 1D substrate with thickness
$d_{\rm sub}$ near an integer multiple of $2\pi c_{\rm light}/(n_{\rm
  sub} B_e)$, where $c_{\rm light}$ is the speed of light and $n_{\rm
  sub}$ is the refractive index of the substrate. This reduces the
reflection coefficient for light waves with perpendicular incidence
(polarized in the $XY$-plane). As a result, the 
important substrate effect comes from the electromagnetic waves
with polarization along $Z$-direction change from the vacuum structure
which can increase the lifetime by a factor of $2$. Another possible
way to increase lifetime can be achieved be use of a glassy substrate
with thickness $d_{\rm sub} \ll B^{-1}_e$ where $B_e$ is given in
cm$^{-1}$. In such cases due to the long wavelength of the resonant
light, the substrate will be invisible. Such a substrate can stand on
thin pillars and as a result any macroscopic object will effectively
be far away from the molecule.

\textbf{\textit{Loss induced by vibrational modes of the substrate --}}
The presence of long-wavelength vibrations in the 2D substrate
also induces vibrations of the nano-rods. This leads to an
phonon-assisted coupling between the molecular rotational 
levels. As a result, there will be transitions between rotational
levels leading to heating (similar to the radiative loss due to
electromagnetic coupling). To gain a qualitative understanding, we
model the surface of the substrate as a square lattice of
atoms. Moreover, we also consider that the underlying arrangement of
atoms in the nano-rods are also cubic. For simplicity, we assume that
atoms in both lattices have mass $M$ and lattice constant $a_S$. We
denote the equilibrium position of individual nano-rod by
$\vec{b}$. 
For long-wavelength phonons, vibrations of the atoms in the nano-rods are
all locked to the 
surface vibrations of the substrate. We first consider the effect of
transverse acoustic phonon modes of the substrate surface. The
displacement of the atoms are normal to the surface with magnitude
$\bm{U}[\vec{b}]$. In second quantized form, we write the displacement
operator in momentum space (two-dimensional momentum
$\vec{k}=(k_x,k_y)$) as  
\begin{eqnarray}\label{phon}
\bm{U}[\vec{b}] &=& \sum_{\vec{k}}\bm{U}_{\vec{k}} e^{i\vec{k}\cdot\vec{b}}, \nonumber\\
\bm{U}_{\vec{k}} &=& \sqrt{\frac{\hbar}{2M \omega_{\vec{k}}r^2_d}} \left(\bm{a}_{\vec{k}}+\bm{a}^{\dagger}_{-\vec{k}}\right),
\end{eqnarray}
where the appearance of $r_d$ is due to our choice of the unit of distance. The phonon creation and annihilation operators are denoted by
$\bm{a}^{\dagger}_{\vec{k}},\bm{a}_{\vec{k}}$ and the phonon
dispersion relation is given in the long wavelength limit as
$\omega_{\vec{k}}= ck $, where $c$ is the sound velocity. In the limit
where the molecules are far away from the nano-rods, the electric
field components due to the transverse displacement of a nano-rod
are expressed as $\bm{E}_{\eta}=\partial E_{\eta}\bm{U}[\vec{b}]$, where
$\eta=\pm,Z$ and $\partial E_\eta = \frac{\partial E_\eta}{\partial Z}$ with the electric field given by Eq.~\eqref{efield-indiv} and as we are using the expression for only a single rod, $m_x=m_y=0$ in this particular case. The total Hamiltonian is given by 
\begin{eqnarray}\label{phham}
H_{\rm tot} &=& H_{\rm rot} + H_{\rm ph} + H_{\rm mol-ph}, \\
H_{\rm rot} &=& \hbar B_e\sum_{\mathcal{N},\mathcal{M_N}} \mathcal{N}(\mathcal{N}+1) \ket{\mathcal{N},\mathcal{M}_\mathcal{N}}\bra{\mathcal{N},\mathcal{M}_\mathcal{N}}, \nonumber\\
H_{\rm ph} &=& \sum_{\vec{k}} (\bm{n}_{\vec{k}}+1/2)\hbar\omega_{\vec{k}}, \nonumber\\
H_{\rm mol-ph} &=& \sum_{\vec{k}}\bm{U}_{\vec{k}} \sum_{\substack{\mathcal{N'},\mathcal{M_N'} \\ \mathcal{N},\mathcal{M_N}}} 
\!\!V_{\mathcal{N'},\mathcal{M}_\mathcal{N'};\mathcal{N},\mathcal{M}_\mathcal{N}}\ket{\mathcal{N'},\mathcal{M}_\mathcal{N'}}\bra{\mathcal{N},\mathcal{M}_\mathcal{N}},\nonumber
\end{eqnarray}
where in order to write the molecule-phonon interaction, we have
assumed that we are interested in the long-wavelength limit and the
molecular matrix element 
$V_{\mathcal{N'},\mathcal{M}_\mathcal{N'};\mathcal{N},\mathcal{M}_\mathcal{N}}=\bra{\mathcal{N},\mathcal{M}_\mathcal{N}} \partial
E_Z \bm{T^1_0} + \partial E_{-}\bm{T^1_1}+\partial E_{+}\bm{T^1_{-1}}
\ket{\mathcal{N'},\mathcal{M}_\mathcal{N'}} $. For our present paper,
we are specifically interested in transition rates from the molecular
states $\ket{\mathcal{N}=1,\mathcal{M}_\mathcal{N}=\pm 1}$ which will
couple to the states $\ket{\mathcal{N}=0,2,\mathcal{M}_\mathcal{N}=0},
\ket{\mathcal{N}=2,\mathcal{M}_\mathcal{N}=\pm 1, \pm 2}$ via
absorption or emission of phonons with energy corresponding to the
energy difference between the molecular levels. Similar to the
electromagnetic case in Ref.~\cite{radelec}, we find
  for the transition rate from $\ket{\mathcal{N}=1,\mathcal{M}_\mathcal{N}=+1}$ to
  $\ket{\mathcal{N},\mathcal{M_N}}$
\begin{align}\label{transition}
\gamma_{\rm ph}[\mathcal{N},\mathcal{M_N}]&= \sum_{\vec{k},\vec{k}'}\sum_{n^i_{\vec{k}}, n^f_{\vec{k}'}}\delta[(\mathcal{N}^2 +\mathcal{N}-2)\hbar B_e + (n^f_{\vec{k}'}-n^i_{\vec{k}})\hbar \omega_{\vec{k}}] \nonumber\\
&\hspace*{-1cm}\times P[n_{\vec{k}}] \left |\bra{n^f_{\vec{k}'}}\bra{\mathcal{N},\mathcal{M_N}} H_{\rm mol-ph} \ket{1,1} \ket{n^i_{\vec{k}'}} \right |^2,
\end{align}
where $\ket{n^{i,f}_{\vec{k}}}$ are phonon Fock states and
$\ket{1,1}=\ket{\mathcal{N}=1,\mathcal{M_N}=1}$. The thermal
distribution of the phonon number $n^i_{\vec{k}}$ is
given by 
$P[n^i_{\vec{k}}]=\exp[-n^i_{\vec{k}}\hbar \omega_{\vec{k}}/(k_B
T)]/(\sum^{\infty}_{m_{\vec{k}}=0} \exp[-m_{\vec{k}}\hbar
\omega_{\vec{k}}/(k_B T)])$ with $T$ being the temperature. The delta
function in Eq.~\eqref{transition} represents the resonance
condition. 

For general $^1\Sigma$ molecules, the rotational energy gap is in the GHz range. For a substrate with sound velocity $c=5\cdot 10^3$m$\cdot$s$^{-1}$, the corresponding phonon wavelength is on the order of $k^{-1}\sim 10^{-6}$m which is much larger than the lattice constant $a_S$. 
Using Eqs.~(\ref{phon}, \ref{elecphon}, \ref{phham}, \ref{transition}), we calculate the transition rates
\begin{equation}\label{phloss}
\begin{array}{lll}
\gamma_{\rm ph}[0,0] &=\frac{\hbar D[\omega_{\vec{k}}]|\partial E_{+}|^2}{3 M r^2_d \omega_{\vec{k}}} \sum_{n}\left(n+1\right) P[n], & \omega_{\vec{k}}=2B_e, \\
\gamma_{\rm ph}[2,\eta] &=f_{\eta}\frac{\hbar D[\omega_{\vec{k}}]|\partial E_{+}|^2}{M r^2_d \omega_{\vec{k}}} \sum_{n} n P[n], & \omega_{\vec{k}}=4B_e,
\end{array}
\end{equation}
where $\eta=0,1,2$, $f_{0}=1/15, f_1=1/5, f_2=2/5$ and 
$D[\omega_{\vec{k}}] = a^2_S\omega_{\vec{k}}/(\pi^2 c^2)$ is the
density of states for the phonon. 
The total transition rate from the
$\ket{\mathcal{N}=1,\mathcal{M}_\mathcal{N}=1}$ state is given by
$\gamma^{\rm tot}_{\rm ph}=\gamma_{\rm ph}[0,0]+\gamma_{\rm
  ph}[2,0]+\gamma_{\rm ph}[2,2]+\gamma_{\rm ph}[2,1]$. For
a quantitative estimate, we assume a sound velocity of $c=5\cdot
10^3$m$\cdot$s$^{-1}$ (similar to the one in a quartz crystal),
temperature $T=10$K, an atom mass for the substrate $M=5 \cdot
10^{-26}$kg (mass of silicon), and a typical lattice constant of
$a_S\sim 0.5$nm. We place a molecule at a distance $\sqrt{X^2+Y^2}
\sim 1, Z\approx 16$ which is similar to the trapping distance in our
scheme with $r_d=60$nm. Inserting the parameters in
Eq.~\eqref{phloss}, for RbCs molecules, the transition rate becomes {$\gamma_{\rm ph}^{\rm tot}\approx 0.01$s$^{-1}$}. To see the combined effect of the four nano-rods for a trapped
molecule near the center of a square cell, we find that the heating rates in Eqs.~\eqref{phloss} are
multiplied by a factor $|\sum_{j_x=\pm 1/2,j_y=\pm 1/2}
(-1)^{j_x+j_y}\exp[-ia_d(k_xj_x+k_yj_y)]|^2$. This is due to the 
the phase in phonon amplitude (Eq.~\eqref{phon}) and the
alternating polarization of the rods. For long phonon
wavelengths $k^{-1}\gg a_d$ this results in a destructive interference
and as a result the transition rate is decreased by a factor $\sim
(ka_d)^4 \sim 10^{-4}$. Hence near the center of the square cell in
Fig.~\ref{figure1}, molecules are more stable than in the corners. 

Phonons with energies in the range of the trapping frequency can further
heat up the molecules by coupling the motional states. Following
a similar procedure as above, we found that the motional heating rate is
dominated by the vibrations along the surface of substrate. The
heating rate is given by  
$$
\gamma_{\rm motion} \propto \omega^2[a_d,Z] \left(\frac{\hbar D[\omega_{\vec{k}}]}{M \sigma^2[a_d,Z] \omega_{\vec{k}}}\right ) \frac{k_B T}{\hbar\omega_{\vec{k}}}, \omega_{\vec{k}}=\omega[a_d,Z]
$$
where the trapping frequency and width are given in
Eqs.~\eqref{osc}. The $\omega^2$ term on the right hand side comes from the square of the overlap of motional states. The second
factor (in parenthesis) originates from the overlap between the
resonant phonon states, whereas the last fraction gives the thermal
phonon number at the resonant frequency. For a molecule trapped at
$Z\sim 1\mu$m ($Z=16$ in units of $r_d$) from top of the nano-rod
with radius $r_d=60$nm, from Eq.~\eqref{osc} we find that $\omega[2.25, 16] \sim
0.5$MHz. For a temperature of $T=4$K this given
a heating rate of $ \gamma_{\rm motion} \lesssim 10^{-2}$s$^{-1}$.

\section{1D nano-traps for molecules}
\label{1dtrap}
Using our $0$D nano-rod arrangement as a building block, we extend to a $1$D structure by repeating the primitive square cell with a lattice constant of $a_{\rm latt}$ as shown in 
Fig.~\ref{figure1}(b). Additionally, the polarization arrangement in each square cell is $\pi/2$ out of phase with its neighbour. Each nano-rod is centered at
$\vec{F}_{q,\bm{m}}=a_{\rm latt}q\hat{X}+a_d(m_x \hat{X}+m_y\hat{Y})/2$,where $q\in[-N_f,-N_f+1,\cdots,-1,0,1,\cdots,N_f-1,N_f]$. The total number of cells is given by
$2N_f+1$. The polarization 
of each rod is defined as, $P[\vec{R}]=(-1)^{q+(m_x+m_y)/2} P \hat{Z}$
when $0<|\vec{R}-\vec{F}_{q,\bm{m}}|<r_d, -h<Z<0$, otherwise it is zero. Also, the $0$D structure can be considered as a special case of $1$D structure with $a_{\rm latt}=\infty$.

In the $1$D structure, we define electric fields equivalent to Eqs.~\eqref{efield-indiv}  by replacing $R_{\bm{m}} \rightarrow R_{q,\bm{m}}$ with $\vec{R}_{q,\bm{m}}=\vec{R}-\vec{F}_{q,\bm{m}}$ and $\phi_{\bm{m}} \rightarrow \phi_{q,\bm{m}}$ with  $\tan{\phi}_{q,\bm{m}}=(Y-m_ya_d)/(X-qa_{\rm latt}-m_xa_d)$.
Similar to Eq.~\eqref{efield}, the total electric field is given by.
\begin{align}
E_{\eta}[\vec{\rho}]&=\sum^{N_f}_{q=-N_f}\sum_{\bm{m}}(-1)^{q+\frac{m_x+m_y}{2}}E_{\eta}[q,\bm{m};\vec{\rho}] \,\,\,(\eta=Z, -).
\end{align}
 
\begin{figure}
    \includegraphics[width=0.45\textwidth]{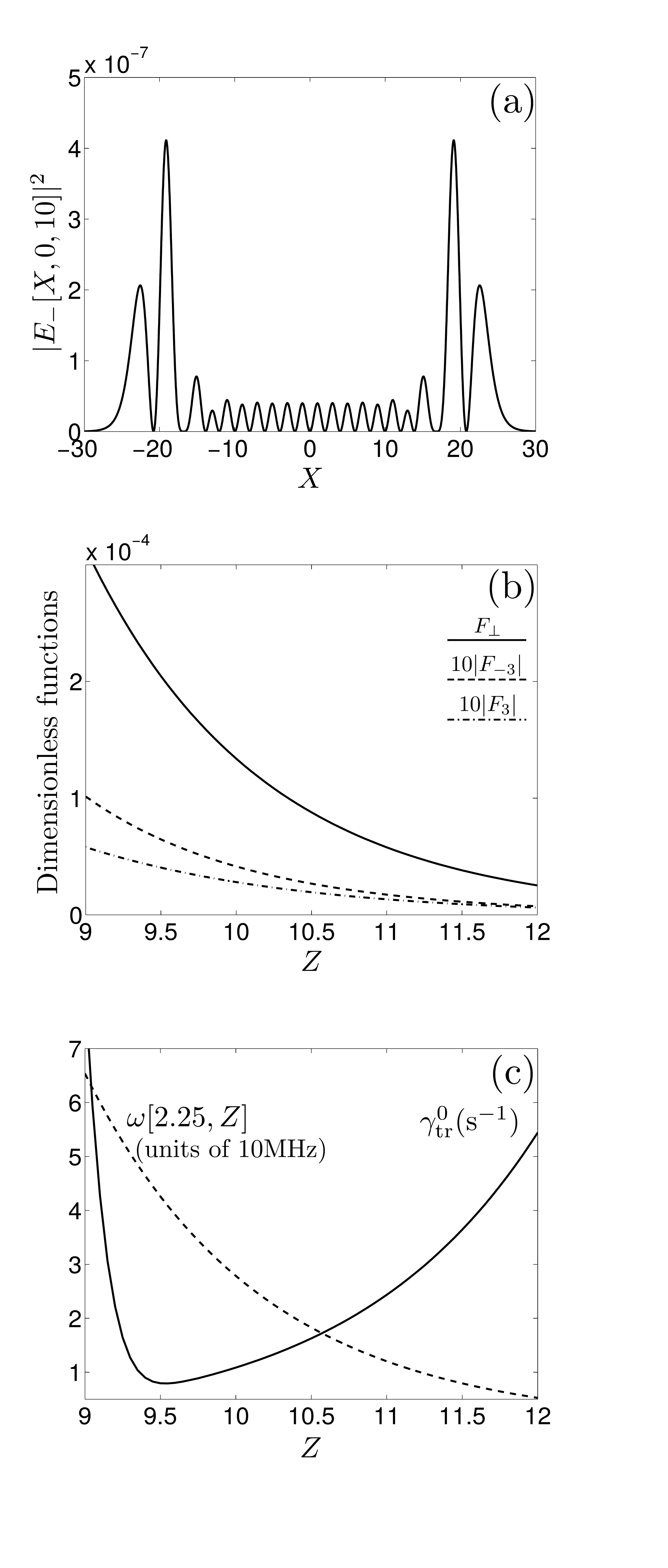}
    \caption{ (a) Shape of the trap potential due to the azimuthal field for a lattice with $N_f=10$, $a_d=2.25$ and $a_{\rm latt} = 2a_d$. (b) Plot of fitting parameters $F_\eta$ defined in
      Eq.~\eqref{fitting2}: $F_{\perp}[2.25,4.5,Z]$ (solid line),
      $10F_{3}[2.25,4.5,Z]$ (dash-dotted line), and $-10F_{-3}[2.25,4.5,Z]$
      (dashed line). (c) The dashed line represents the  
      zero-point trap frequency $\omega[2.25,4.5,Z]/2$ in units of MHz. The solid
      line is $\gamma^0_{\rm tr}$ , the total molecule loss rate 
for the $j=0$ hyperfine state (in Hz).}  
        \label{figure3a}
\end{figure}

To look for the properties of the electric field, we first notice that the long-range nature of the dipole potential from
  neighboring cells strongly affects the trapping potential and lowers it significantly. Moreover, the absence of rotational
  symmetry leads to additional loss terms. 
Each square cell is centered at $(q,0)a_{\rm latt}$ and is bounded by
the lines $X=(qa_{\rm latt}\pm a_d/2) ,|Y|=a_d/2$ as shown in
Fig.~\ref{figure1}(b). The alternating orientation of the cells generate electric-field
distributions with out of phase neighboring square cells. Inside each square cell, the trap potential {close to
the center}
depends quadratically on $|E_{-}|$, same as in $0$D. In
Fig.~\ref{figure3a}(a), we plot the $|E_{-}|^2$ along the $X$-axis for
a fixed $Z$ with $N_f=10, a_{\rm latt}=2a_d$. It is clear that there
is a potential minimum at the center of each square cell whereas there
is a shift in the position of the  minima for the boundary square
cells. Moreover, the trap height at the boundary is higher which will
result in a reduced escape rate of the molecules from the boundary traps.     

For a quantitative study, we define a local polar coordinate at each cell $q$ as, $R^2_q=(X-qa_{\rm latt})^2+Y^2$ and $\varphi_q=\tan[\frac{Y}{X-qa_{\rm latt}}]$. Similar to the $0$D trap, the 1D trap is symmetric under reflection at the $X$- or $Y$-axes (and
change of polarization) and the Fourier coefficients of the azimuthal field $E_{-}$ with even power vanishes. The rotational symmetry about center of
the square is violated. However, as noted in Appendix~\ref{1dapp}, the leading
order term in $E_{-}$ is still $\propto iRe^{i\varphi}$. Thus we can
write the electric field {at the center of the cell} as 
\begin{eqnarray}\label{fitting2}
E_{-q}[\vec{\rho}] &\approx& iR_q
(F_{\perp}[a_d,a_{\rm latt},Z])\exp[i \varphi_q]\nonumber\\
&+& i\sum_{\eta=\pm 3}F_{\eta}[a_d,a_{\rm latt},Z]R^3_q\exp[i\eta\varphi_q], \nonumber\\
E_{Zq}[\vec{R}_q,Z] &\approx& F_{z}[a_d,a_{\rm latt},Z]R^2_q\sin [2\varphi_q],
\end{eqnarray}
where $R_q<a_d/2$. In contrast with the $0$D case in
Eq.~\eqref{series0d}, the electric field contains both of $\pm 3$
azimuthal components. The fitting functions are 
 %---------------------------
%----------------------------
shown in Fig.~\ref{figure3a}(b) for parameters in Table.~\ref{tab:table1}. Comparing with the $0D$ case
(Fig.~\ref{figure2a}) we see that for similar $a_d$,
$|F_{\perp}[a_d,a_{\rm latt},Z]| < |f_{\perp}[a_d,Z]|$ and as a result yields much weaker potential.  Only in the situation of $a_{\rm latt} \rightarrow \infty$, they become equal as the $1$D trap becomes equivalent to $0$D trap. Otherwise, magnitude of $|F_{\perp}[a_d,a_{\rm latt},Z]|$ can be boosted by trapping the molecules nearer to the nano-rods and by increasing the total polarization of the nano-rods. Moreover, for fitting functions to $R^3$ component (responsible for non-adiabatic loss), compared to the $0$D case, $\frac{|F_{\pm 3}[a_d,a_{\rm latt},Z]|}{|F_{\perp}[a_d,a_{\rm latt},Z]|} \gtrsim \frac{|f_{-3}[a_d,Z]|}{|f_{\perp}[a_d,Z]|}$ which results in larger non-adiabatic loss rates.

For the $1$D potential using the fields in
Eq.~\eqref{fitting2} we solve the equivalent of equation
Eq.~\eqref{eigeneq21} within the cell
$q$. Following Eq.~\eqref{oscen}, the lowest energy trapped state is
\begin{align}\label{1dstate}
\ket{t^j;\ell;N;\mathcal{N}=1}_q &= \ket{t_q,\ell N} \ket{\mathcal{N}=1,-,\varphi_q}, \nonumber\\
\bra{\vec{\tilde{R}}_q\mid t_q,\ell N} &= \frac{\sqrt{2}\left(\tilde{R}_q\right)^{\ell_{\rm eff}}\exp\left[-\tilde{R}^2_q/2\right]\mathcal{L}^{\ell_{\rm eff}}_N\left[\tilde{R}^2_q\right]}{(\Gamma[N+\ell_{\rm eff}+1]/N!)^{1/2}}, 
\end{align}
where $\vec{\tilde{R}}_q=\vec{\tilde{R}}-q\tilde{a}_{\rm
  latt}\hat{X}$, with
$\tilde{a}_{\rm latt}={a}_{\rm latt}/\sigma[a_d,{a}_{\rm latt},Z]$, and the $Z$-dependent oscillator
width and frequency have the form
\begin{eqnarray}\label{1dosc}
\sigma^2[a_d,{a}_{\rm latt},Z]&=&\frac{\left(10 K \hbar
  B_e\right)^{1/2}}{F_{\perp}[a_d,{a}_{\rm latt},Z] \alpha_{\rm mf}}, \nonumber\\ 
\hbar\omega[a_d,{a}_{\rm latt},Z] &=& 
|\alpha_{\rm mf}|F_{\perp}[a_d,{a}_{\rm latt},Z] \left(\frac{2K}{5\hbar B_e}\right)^{1/2}.
\end{eqnarray}
We plot the trap frequency for the parameters in Tables~\ref{tab:table1}, \ref{tab:table2} in Fig.~\ref{figure3a}(c) (the dashed line). 

{\it $Z$-trapping in $1$D}: To prevent the molecule from escaping in
$Z$-direction, we use the same laser set-up as discussed in
Sec.~\ref{lassec}. As an example, from Fig.~\ref{figure3a}(c), an optimum
position to trap the molecule will be around $Z=9.5$ where the trap frequency
$\omega[2.25,4.5,9.5] \approx 10$MHz. To look for laser parameters, we use
Eq.~(\ref{effp}) as an expression for an effective potential along $Z$-axis
with $Z_1 \approx 9.5$. This is fulfilled for a total laser power
$I_0=1.0$MW$\cdot$cm$^{-2}$ and the focal plane $Z_0=7.65$. The corresponding
$Z$-axis trap frequency $\omega_Z=0.5$MHz and oscillator length
$\sigma_Z=0.2$. Though the $Z$-trapping is much weaker than the radial
trapping, the oscillator length is still less than the nano-rod
radius. One important
change from the $0$D case is the laser induced loss-rate which will be around
$\approx 1$s.

{\it Molecular loss rate:} Now, we estimate the molecular loss rate
for $1$D by following similar procedures as discussed in
Sec.~\ref{nonad} with the fitting functions denoted by $f$ is replaced
by the corresponding $1$D functions, $F$. For each cell, we calculate the $1$D equivalent of the molecular loss rate following the treatment for Eqs.~(\ref{decaytrapr4}, \ref{decaycontr4}), the total molecule loss rate for the $j=0$ state ($\gamma^0_{\rm tr}$) is shown in Fig.~\ref{figure3a}(c).
For the $1$D case, the calculation of the non-adiabatic loss rates due
to the $R^4$ potential gives results similar to
Eqs.~(\ref{decaytrapr4}, \ref{decaycontr4}), with $F_{\pm 3}$ in place
of $f_{-3}$. Again we notice the interplay between non-adiabatic loss
($\propto \omega^{-1}[a_d,a_{\rm latt},Z]$) and hyperfine-induced molecule
loss ($\propto \omega[a_d,a_{\rm latt},Z]$). Due to the relatively large non-adiabatic coupling (compared to $0$D), controlled by the ratio $\frac{|F_{\pm 3}[a_d,a_{\rm latt},Z]|}{|F_{\perp}[a_d,a_{\rm latt},Z]|}$, 
we need a high trap frequency to minimize the non-adiabatic loss rate.
The hyperfine-induced loss rate is controlled by the potential barrier between the $j=0,1$ hyperfine states: $|E_0-E_1|/\omega[a_d,a_{\rm latt},Z]$, which in turn is controlled by the magnetic field. Accordingly, we need a stronger magnetic field, see Table ~\ref{tab:table3}, to increase the potential barrier between the hyperfine states. The minimum loss rate we obtain for the parameters in Table ~\ref{tab:table3}: $\gamma^0_{\rm tr} \approx 0.8$s$^{-1}$ for $Z\approx 9.5$. One way to increase lifetime will be by increasing $a_d$ or $a_{\rm latt}$ which will result in increased lattice constant and lower the energy scales. 

Moreover, as we are in a lattice, we like to have a stable local trap,
i.e., we want to have a small tunneling rate, which is guaranteed as long as $a_{\rm latt}/\sigma \gg 1$ \cite{dutta}. From Eq.~\eqref{1dosc}, we see that the ratio
$a_d/\sigma \sim |F_{\perp}[a_d,a_{\rm latt},Z]| a_{\rm mf}$, where $F_{\perp}[a_d,a_{\rm latt},Z]$ is weaker than the corresponding $0$D fitting function
$f_{\perp}[a_d,a_{\rm latt},Z]$ for fixed $a_d,a_{\rm latt}, Z$. As a result, we need to increase the ferroelectric
strength $\alpha_{\rm mf}$ to get the same $a_{\rm latt}/\sigma[a_d,a_{\rm latt}, Z]$. Therefore we chose in
Table~\ref{tab:table1} the polarization strength for $1$D 
stronger than that of $0$D. Consequently, one needs to use ferroelectric
material with high spontaneous polarization like PZT (Lead zirconate
  titanate compounds). For such parameters we find that
$a_{\rm latt}/\sigma[2.25,5,9.5] \sim 10^2$, which implies an effectively vanishing tunneling rate. 

Next we consider the substrate-induced loss rate as discussed in
Sec.~\ref{sec:surfloss}. Compared to the $0$D case, the trap center is
closer to the surface at $Z\approx 9.5$. As a result, for the parameters
in Table~\ref{tab:table1}, we obtain from Eq.~\eqref{eq:1} a loss rate
$\gamma_h \sim 0.2$s$^{-1}$. We also obtain a similar loss rate for phonon induced noise from Eqs.~\eqref{phloss}. From these discussion it is clear that the most important loss mechanism originates from non-adiabatic and hyperfine coupling with molecule lifetime on the order of $1$s (Fig.~\eqref{figure3a}(c)).

\section{Simulation of quantum spin model:}
\label{qsim}
From the discussion in the previous section it is clear that each
primitive cell of the ferroelectric lattice can trap molecules. For the
$\mathcal{N}=1$ manifold the trapped state at a site $q$ is given by
Eq.~\eqref{1dstate}. Applying a laser field similar
to Sec.~\ref{lassec}, we trap the molecule at $Z=Z_0$.

To use the molecules as spins, we need a second trapped state. We have
carried out similar studies as in the previous sections for the level
$\mathcal{N}=2$. Let $\ket{\pm1}$ again denote the $\mathcal{M_N}=\pm1$
projection on the molecular axis. The quadratic Stark shift for
$\mathcal{N}=2$ level is weaker than for $\mathcal{N}=1$. Numerically, we
find that the effective potential (Eq.~\eqref{trapham})
is $V^{\mathcal{N}=2}_{\rm rad}\approx V^{\mathcal{N}=1}_{\rm rad}/4$. We
then solve the equivalent of Eqs.~(\ref{eigeneq21}, \ref{eigeneq22}, \ref{eqZ}), only changing the potential 
strength. We denote the transverse trapping width for $\mathcal{N}$ manifold as $\sigma_{\mathcal{N}}[a_d,Z]$. In relation to the oscillator
width and energy for the $\mathcal{N}=1$ state, $\sigma_2[a_d,a_{\rm latt},Z]\approx
\sqrt{2}\sigma_1[a_d,a_{\rm latt},Z]$ and
$\omega_{2}[a_d,a_{\rm latt},Z]=\omega_1[a_d,a_{\rm latt},Z]/2$. For the laser induced
potential, as noted in Ref.~\cite{jin}, the polarizability of the
molecule is almost independent of $\mathcal{N}$ and as a result we
trap the $\mathcal{N}=2$ state also at $Z_0$. 
Following Eq.~\eqref{1dstate}, the corresponding trapped state is then expressed as
\begin{align}\label{2dstate}
& \ket{t^j;\ell;N;\mathcal{N}=2}_q = \ket{t'_q,\ell N} \ket{\mathcal{N}=2,-,\varphi_q}, \nonumber\\
& \braket{\vec{\tilde{R}}_q \mid t'_q,\ell N} = \frac{\sqrt{2}\left(\tilde{R}_q\right)^{\ell_{\rm eff}}\exp\left[-\tilde{R}^2_q/4\right]\mathcal{L}^{\ell_{\rm eff}}_N\left[\tilde{R}^2_q/2\right]}{2^{\ell_{\rm eff}/2}(\Gamma[N+\ell_{\rm eff}+1]/N!)^{1/2}}, 
\end{align}
For this section, we only consider the motional states $\ell=0, N=0$ and as a
result omit these labels in our description of the states. To simulate a spin-model with long-range dipolar interaction, we first consider just two cells at $q$ and $q'$ and assume that each cell is filled with one molecule in the state $\ket{t^0;\mathcal{N}=1}$ or $\ket{t^0;\mathcal{N}=2}$. We introduce the spin operators,
\begin{eqnarray}
\bm{S}^{+}_q&=&\ket{t^0;1;0;\mathcal{N}=2}_q\bra{t^0;1;0;\mathcal{N}=1}_q, \nonumber\\
\bm{S}^{z}_q&=&\sum_{\eta=1,2}(-1)^{\eta}\ket{t^0;1;0;\mathcal{N}=\eta}_q\bra{t^0;1;0;\mathcal{N}=\eta}_q, \nonumber\\
\end{eqnarray} 
and $\bm{S}^{-}_q=[\bm{S}^{+}]^{\dagger}_q$. The {dipole-dipole}
Hamiltonian {projected to the subspace of interest} is then given by,
\begin{align}\label{hamdip}
H_{\rm class} &=& V_{\rm dd}\sum_{q\neq q'}\frac{\bm{S}^{+}_q\bm{S}^{-}_{q'}+(\bm{S}^{+}_q\bm{S}^{+}_{q'}+\bm{S}^{-}_q\bm{S}^{-}_{q'})/2}{|q-q'|^3}+4\hbar B_e\sum_q\bm{S}^z_q, \nonumber\\
\end{align}
where dipolar energy is given by
\begin{align}
  \label{eq:Vdip}
  V_{\rm dd} & =\frac{\mu^2\sum^{}_{\mathcal{M_N}=\pm 1}|\braket{\mathcal{N}=1,\mathcal{M_N} \mid \bm{T^1_0}\mid \mathcal{N}=2,\mathcal{M_N}}|^2}{8\pi\epsilon_0a^3_{\rm latt}}, \nonumber\\
             & =  \frac{1}{5}\frac{\mu^2}{4\pi\epsilon_0a^3_{\rm latt}}
\end{align}
with the factor of $1/5$ originating from the dipole matrix between
the $\mathcal{N}=1$ and $\mathcal{N}=2$ states and the last term denotes detuning between the two spin states. For the parameters concerned, $V_{\rm dd} \ll \hbar B_e$, $\bm{S}^z_q$ becomes a conserved quantity and as a result, the pair creation and annihilation terms $\bm{S}^{+}_q\bm{S}^{+}_{q'},\bm{S}^{-}_q\bm{S}^{-}_{q'}$ in Eq.~\eqref{eq:Vdip} are suppressed. Hence, one has a long-range classical Ising model.

To simulate a quantum model, one way is to couple the $\ket{t^0;1;0;\mathcal{N}=2}_q, \ket{t^0;1;0;\mathcal{N}=1}_q$ state by introducing a linearly $Z$-polarized time-periodic microwave field $\vec{E}_{\rm mw}=\mathcal{E} \cos{\Omega t}\hat{Z}$ where
$\Omega=4B_e+\Delta$. The microwave coupling Hamiltonian is given by,
$H_{\rm mv}=g_{\rm mw}\cos{\Omega t}\sum_q\bm{S}^x_q$, where $g_{\rm mw}=\frac{\mu \mathcal{E}}{\sqrt{5}} \braket{t'_q,1 0\mid t_q,10}$.
Going to the rotating frame and projecting to
the trapped state, the spin Hamiltonian has the form (see Appendix \ref{app:hsim}),
\begin{align}\label{hsim}
H_{\rm spin} &= V_{\rm dd}\sum_{q,q'}\frac{\bm{S}^{+}_q\bm{S}^{-}_{q'}}{|q-q'|^3}+\hbar(\Delta-\omega_1[a_d,Z_0]/2)\sum_q\bm{S}^z_q \nonumber\\
&+ g_{\rm mw}\sum_q\bm{S}^x_q,
\end{align}
The second term {in
Eq.~\eqref{hsim}} originates from the detuning of the microwave field
and the difference in trap frequency of the two trapped states. 
The microwave-molecule coupling gives the last term,
where we have assumed that the width of the wave-function in $Z$
remains same in both levels. The Hamiltonian in Eq.~\eqref{hsim} is an
example of a long-range $XX$ spin Hamiltonian in a transverse and
longitudinal field, both of which are tunable. 

Note, however, that the presence of more than one molecule and the
dipolar interactions also lead to a new loss mechanism: dipolar collisions
between two molecules can also resonantly couple the states
$\ket{t;\mathcal{N}=2}_q\ket{t;\mathcal{N}=1}_{q'}$
to the untrapped states $\ket{\mathcal{N}=2,\mathcal{M_N}=0}
\ket{\mathcal{N}=1,\mathcal{M_N}=0}$, where $q,q'$ are two sites
in the lattice. As discussed in Sec.~\ref{lassec}, the
$\mathcal{M}_{\mathcal{N}}=0$ state is detuned from the
$\mathcal{M}_{\mathcal{N}}=\pm 1$ state. As a result, the loss rate is
suppressed by a factor (see App.~\ref{app:hsim}),  
\begin{equation}
  \label{eq:diploss}
  (V^2_{\rm
    dd}/K)\exp\left[-2\frac{\alpha_0I_0}{K}\right].
\end{equation}
For a laser strength of $I_0 \sim 1.0$MW$\cdot$cm$^{-2}$,
and the polarizability was taken from Ref.~\cite{kot1}, and using parameters from Table \ref{tab:table1}, we find that $\alpha_0I_0/K \sim 10^2$. We see that such a loss
rate is exponentially suppressed.

Moreover, to suppress motional excitation, one needs to
make sure that the dipolar energy remains much weaker than the local
trap energy: $V_{\rm dip}<2\hbar\omega[a_d,a_{\rm latt}, Z]$. This gives
a lower limit on the lattice constant $a_{\rm latt}$ for a fixed molecule and
ferroelectric polarization. For the parameters in
Tables~\ref{tab:table1} and
\ref{tab:table2} with a molecule trapped at $Z_0=9.5$ (minimum loss rate from Fig.~\ref{figure3a}(c)), we find that the dipolar energy $V_{\rm dip} \approx 6$kHz which is small compared to the trap energy of $\sim 1$MHz. 
Moreover, as our trap is stable for around $\sim 1$s, such a dipolar
coupling in principle allows to perform around $V_{\rm dip}/\mathrm{min}[\gamma^0_{\rm tr}] \approx 7\cdot10^3$ gate
operations. If we use instead a $5$T magnetic field to suppress the hyperfine
loss, we can gain an order of magnitude in the number of gate operations by
decreasing the nano-rod radius $r_d=20$nm. For a comparison to a possible optical
lattice trap, we assume a setup similar to Ref.~\cite{nagerl} with an
additional microwave field to couple $\ket{\mathcal{N}=0,\mathcal{M_N}=0}$ to
$\ket{\mathcal{N}=1,\mathcal{M_N}=0}$. The resulting spin model permits gate
operation of order $\sim 3\cdot 10^2$. Thus, in our setup we can expect an
increase in the number of potential gate operations by a 
factor of $\sim 20$.

\section{Summary and Conclusions}

In summary, we have proposed nano-traps for polar molecules near an
array of ferroelectric nano-rods. Our most important finding is that
in the proposed scheme there exist trapped states with suppressed
molecular loss rate even within the regime of nano-scale
confinement. The molecules are held at a certain distance from the
nano-rods by {the combined potentials of the nano-rods and a
  standing-wave laser field}. Moreover, we have shown that the
trapping scheme can be extended to an one dimensional periodic
structure with lattice constant $\sim 200$nm. We carried out a
qualitative analysis of the main loss mechanisms, including
non-adiabatic losses as well as hyperfine-, laser- and surface-induced
losses. {Considering, in particular, RbCs molecules that have
  already been prepared at temperatures below those corresponding to
  our trap, we} find that the main limiting process comes from the
interplay between non-adiabatic and hyperfine coupling and leads to a
lifetime of $\sim 10$s for $0$D trap and $\sim 1$s for $1$D trap. This
time, in principle, can be increased by applying a stronger static
magnetic field. As a potential application, described a way to
simulate a family of long-range spin Hamiltonians using our proposed
traps. In principle, one can also reach quasi two dimensional regime by using a stack of $1$D traps presented here.

We like to point out that the present proposal can be applied to any
$^{1}\Sigma$ diatomic molecule. Depending on the hyperfine and
rotational structure of the molecule, the dependence of the loss rate on
the magnetic field will change. Moreover, one can exploit the
possibility that for $\mathcal{N}>1$, there can be more than one
trapped state which can lead to more exotic spin models. In addition,
difference in trapping properties between different $\mathcal{N}$
manifolds can be exploited to another variety generate spin models by
shaking the $Z$-axis optical potential without using the microwave
field.

We believe that the present proposal opens up a new direction in the
hybrid systems of ferroelectric nano-structures and polar
molecules. One possible extension is to study the trapping of
$^{2}\Sigma$ molecules. Furthermore, ferroelectrics in the nano-regime
can have exotic polarization distribution \cite{vortex, vortex2} which
can be controlled in dynamical manner. Such control can potentially
give rise to state-independent trap in a time-averaged
potential. Another direction will be to extend our trapping scheme to
open shell molecules, e.g., $^{2}\Sigma$ molecules. Such molecules can
then be trapped by ferro\emph{magnetic} nano-rods \cite{magrod}. Moreover,
ferromagnetic states can be switched in nanosecond rates
\cite{magvortex} which can give rise to a novel mechanism to trap
open-shell molecules. 

Another route of further investigation will focus on the usability of
such traps for quantum information processing and for precision
measurements.

\appendix

\section{Symmetries of the trapping potential}
\label{app:symmetry}
 We consider the simplest ferroelectric trapping geometry consisting of
only $4$ nano-rods, i.e., $N_f=1$, corresponding to just one square in
Fig.~\ref{figure1} as outlined by the dashed line. 
The symmetry of the arrangement of ferroelectrics gives useful insight
into the properties of the electric field. 
There are three relevant symmetries $R_J, J=1,\dots3$ involved:
reflection at $Y=0$, reflection at $X=0$, and rotation around $Z$ by
$\pi/2$. Each of these operations moves a $\pm$ polarized nano-rod to
a $\mp$ polarized one. Therefore, the electric field  $\vec{E}(\vec{R})$
changes sign under each of the symmetry transformations $R_J$, i.e. \cite{Joan2008},
\begin{equation}
  \label{eq:3}
  R_J\vec{E}(R_J^{-1}\vec{R}) = -\vec{E}(\vec{R})\,\,\forall J=1,\dots3.
\end{equation}
We are interested at the field distribution close to the origin (the
center between the rods) but outside of the volume containing the
nano-rods. Thus we can use a Taylor series in $X,Y$ for each component
of  $\vec{E}$
\begin{equation}
  \label{eq:2}
  E_u(X,Y,Z) = \sum_{n,m\geq0} c^u_{nm}(Z)X^nY^m,\,\, u=X,Y,Z.
\end{equation}
Then from $R_1$ (reflection at $Y=0$), we obtain
\[
\left( \begin{array}{c}
-E_X(-X,Y,Z)\\
E_Y(-X,Y,Z)\\
E_Z(-X,Y,Z)
\end{array} \right)
= -\vec{E}(X,Y,Z), 
\]
which implies:
$c^X_{nm}=(-1)^nc^X_{nm}$ and $c^Y_{nm}=-(-1)^nc^Y_{nm}$, i.e., 
\begin{equation}
  \label{eq:4a}
\begin{array}{lll}
c^X_{nm} =0\, \forall n\,\mbox{odd},&
c^Y_{nm} =0\, \forall n\,\mbox{even},&
c^Z_{nm} =0\, \forall n\,\mbox{even}.
\end{array}
\end{equation}
Similarly, reflection at $X=0$ yields
\begin{equation}
  \label{eq:4b}
\begin{array}{lll}
c^X_{nm} =0\, \forall m\,\mbox{even}, &
c^Y_{nm} =0\, \forall m\,\mbox{odd},&
c^Z_{nm} =0\, \forall m\,\mbox{even}.
\end{array}
\end{equation}
Thus we can conclude that the only non-zero terms in the Taylor series
are $c^X_{\rm odd,even}, c^Y_{\rm even,odd}$, and
$c^Z_{\rm even,even}$.
Note that these symmetries persist even in the 1D case and thus still apply
to the (center of) the 1D array of  nano-rods. $R_3$, in contrast,
only holds for the 0D case (and would be restored for the (center of)
a full 2D arrangement that we do not discuss here).
\begin{align}
  \label{eq:5}
  c^X_{nm} &=-(-1)^mc^Y_{mn},\\
  c^Y_{nm} &=+(-1)^mc^X_{mn},\\
  c^Z_{nm} &=-(-1)^mc^Z_{mn}.
\end{align}
We are mainly interested in the \emph{azimuthal} field component $E_-= 
E_X-iE_Y$. Inserting the Taylor series for $E_{X,Y}$ and expressing
$X$ and $Y$ in polar coordinates, we find
\begin{align*}
  E_-(R,\varphi,Z) &= \sum_{n,m} \left[ c^X_{nm}(Z)-ic^Y_{nm}(Z) \right]
  R^{n+m}\cos^n\varphi\sin^m\varphi.
\end{align*}
Using that both $c^X_{nm}$ and $c^Y_{nm}$ vanish whenever $n+m$ is
even, we see that only odd powers of $R$ appear in the
series and only powers of $e^{il\phi}$ with $|l|<2k1$ appear. Therefore, we are justified to make the ansatz
\begin{equation}
  \label{eq:6}
  E_-(R,\varphi,Z) = \sum_{k\geq0, 2k+1\geq |l| } c^-_{kl}(Z)
  R^{2k+1}e^{il\varphi}\equiv\sum_lc^-_l(Z,R)e^{il\varphi}, 
\end{equation}
where the $c^-_l(Z,R)$ are odd functions of $R$. While further
constraints on $l$ can be obtained (after some algebra) 
by relating $c^X,c^Y$ and $c^-$, they are easier to see by
applying the symmetry operations directly to the Fourier series of 
$E_-$ and using that from the three symmetry operations, we get:
% \begin{align}
%   \label{eq:7a}
%   E_-\left({\tiny \vek{X}{Y}{Z}}\right) &=E_-^*\left({\tiny \vek{-X}{Y}{Z}}\right) = -E_-^*\left({\tiny \vek{X}{-Y}{Z}}\right) = iE_-\left({\tiny \vek{Y}{-X}{Z}}\right)\\
% &= E_-^*({\tiny \vek{R}{\pi-\varphi}{Z}}) = -E_-^*({\tiny \vek{R}{-\varphi}{Z}}) = iE_-({\tiny \vek{R}{\varphi-\pi/2}{Z}}).\\
% &\hspace*{-1cm}= E_-^*(R,\pi-\varphi,Z) = -E_-^*(R,-\varphi,Z) = iE_-(R,\varphi-\pi/2,Z).
% \end{align}
% \begin{align}
%   \label{eq:7b}
%   E_-(X,Y,Z) &=E_-^*(-X,Y,Z) = -E_-^*(X,-Y,Z) = iE_-(Y,-X,Z)\\
% &= E_-^*(R,\pi-\varphi,Z) = -E_-^*(R,-\varphi,Z) = iE_-(R,\varphi-\pi/2,Z).
% \end{align}
%% formatting of these two is ugly
\begin{align}
  \label{eq:7}
  E_-(X,Y,Z) &=E_-^*(-X,Y,Z) \equiv E_-^*(R,\pi-\varphi,Z), \\
& = -E_-^*(X,-Y,Z) \equiv -E_-^*(R,-\varphi,Z),\\
& = iE_-(Y,-X,Z)\equiv iE_-(R,\varphi-\pi/2,Z).
\end{align}
This implies that
$c_l^-=(c_l^-)^*e^{il\pi}=(c_l^-)^*e^{i\pi}=c_l^-e^{-i(m-1)\pi/2}$
from which we conclude that 
\begin{align}
  \label{eq:8}
  c^-_{l} &= 0\, \forall l\not= 4J+1,\\
c^-_{l} &= -(c^-_{l})^*,
\end{align}
i.e., only $c_{l}^-$ with $l=\dots,-3,1,5,\dots$ may be non-zero (neither
even powers of $e^{i\varphi}$ in the Fourier series of $E_-$ nor powers
$4k-1$). \\
Analogously, we find for the Fourier series of
$E_Z(\varphi,R,Z)=\sum_l c^Z_l(R,Z)e^{il\varphi}$ that
$-c^Z_l=e^{il\pi}c_{-l}=c^Z_{-l}=e^{il\pi/2}c^Z_l$ from which we
conclude that $c^Z_l$ may be non-zero only for $l=4J+2$ and that
$c^Z_{-l}=-c^Z_l$; all $c^Z_l$ are \emph{even} functions of $R$.\\
Note that these considerations only apply to $0$D case or the center
of a $1$D or $2$D array. However, due to the diminishing influence of
the boundary, it will approximately hold also for cells close to the
center of such an array.

\section{Non-adabatic contribution to potential Eq.~\eqref{effpot1}}\label{app1}

Here we consider the non-adiabatic effect to the Hamiltonian due to the position dependence of the perturbation in Eq. \eqref{effpot1}. The second-order energy-corrections arise from a first order correction to the unperturbed states. The transformed states related to our model are given by (we neglect the hyperfine structure for this discussion)
\begin{widetext}
\begin{eqnarray}
\ket{\tilde{0},\tilde{0}} &=& \ket{\mathcal{N}=0,\mathcal{M_N}=0} - \sum_{\substack{\mathcal{N'}>0,\\ \mathcal{M_{N'}}}} \frac{\bra{\mathcal{N'},\mathcal{M_{N'}}}H_{\rm mf}\ket{\mathcal{N}=0,\mathcal{M_N}=0}}{\hbar B_e \mathcal{N'}(\mathcal{N'}+1)} \ket{\mathcal{N'},\mathcal{M_{N'}}}, \nonumber\\
&=& \ket{\mathcal{N}=0,\mathcal{M_N}=0} + \frac{\alpha_{\rm mf}}{2\sqrt{6}\hbar B_e} \left( E_{-}\ket{\mathcal{N}=1,\mathcal{M_N}=1} + E_{+}\ket{\mathcal{N}=1,\mathcal{M_N}=-1}\right), \nonumber\\
\ket{{\tilde{1}}}&=& \ket{\mathcal{N}=1,\mathcal{M_N}= 1} + \sum_{\substack{\mathcal{N'}\neq 1,\\ \mathcal{M_{N'}}}} \frac{\bra{\mathcal{N'},\mathcal{M_{N'}}}H_{\rm mf}\ket{\mathcal{N}=1,\mathcal{M_N}=1}}{\hbar B_e (2-\mathcal{N'}^2-\mathcal{N'})} \ket{\mathcal{N'},\mathcal{M_{N'}}}, \nonumber\\
&=& \ket{\mathcal{N}=1,\mathcal{M_N}=1} - \frac{\alpha_{\rm mf}}{4\sqrt{5}\hbar B_e} \left( E_{-}\ket{\mathcal{N}=2,\mathcal{M_N}=2} + \frac{E_{+}}{\sqrt{6}}\ket{\mathcal{N}=2,\mathcal{M_N}=0} - \sqrt{\frac{10}{3}}E_{+}\ket{\mathcal{N}=0,\mathcal{M_N}=0} \right ), \nonumber\\
\ket{{-\tilde{1}}}&=& \ket{\mathcal{N}=1,\mathcal{M_N}=-1} + \sum_{\substack{\mathcal{N'}\neq 1,\\ \mathcal{M_{N'}}}} \frac{\bra{\mathcal{N'},\mathcal{M_{N'}}}H_{\rm mf}\ket{\mathcal{N}=1,\mathcal{M_N}=-1}}{\hbar B_e (2-\mathcal{N'}^2-\mathcal{N'})} \ket{\mathcal{N'},\mathcal{M_{N'}}}, \nonumber\\
&=& \ket{\mathcal{N}=1,\mathcal{M_N}=-1} - \frac{\alpha_{\rm mf}}{4\sqrt{5}\hbar B_e} \left( E_{+}\ket{\mathcal{N}=2,\mathcal{M_N}=-2} + \frac{E_{-}}{\sqrt{6}}\ket{\mathcal{N}=2,\mathcal{M_N}=0} - \sqrt{\frac{10}{3}}E_{-}\ket{\mathcal{N}=0,\mathcal{M_N}=0}\right). \nonumber\\
\end{eqnarray}
\end{widetext}

The non-adiabatic effect then can be estimated by employing the kinetic energy operator onto the states $\ket{\tilde{0},\tilde{0}}, \ket{\pm \tilde{1}}$. As we have seen that the width of the trapped states are within the linear region of the electric field stregth, we only consider the effects of $E_{\pm }$ fields. The resulting kinetic operator reads, $K_{\rm tot}=K_0+K_{1}+K_{2}$ where
\begin{widetext}
\begin{eqnarray} \label{nonad1}
K_0 &=& K \left(\ket{\tilde{0},\tilde{0}}\bra{\tilde{0},\tilde{0}}+\ket{ \tilde{1}}\bra{\tilde{1}}+\ket{-\tilde{1}}\bra{-\tilde{1}}\right) \nabla^2, \nonumber\\
K_{1} &=& 
K \left(\frac{\alpha_{\rm mf} }{2\sqrt{6}\hbar B_e}\right)^2 \sum_{\sigma=\pm}\left(E_{\sigma} (\nabla E_{\sigma})\cdot\left(\nabla \ket{\tilde{0},\tilde{0}}\right)\bra{\sigma\tilde{1}}
+ E_{-\sigma} (\nabla E_{-\sigma})\cdot\left(\nabla\ \ket{\sigma\tilde{1}}\right)\bra{\tilde{0},\tilde{0}} + E_{-\sigma} (\nabla E_{\sigma}) \cdot\left(\nabla \ket{\tilde{0},\tilde{0}}\right)\bra{\tilde{0},\tilde{0}}\right), \nonumber\\
K_{2} &=& K \left(\frac{\alpha_{\rm mf} }{4\sqrt{30}\hbar B_e}\right)^2 \sum_{\sigma=\pm} \left(E_{-\sigma} (\nabla E_{-\sigma})\cdot\left(\nabla \ket{\sigma\tilde{1}}\right)\bra{-\sigma\tilde{1}}+ E_{\sigma} (\nabla E_{-\sigma})\cdot\left(\nabla \ket{\sigma\tilde{1}}\right)\bra{\sigma\tilde{1}}\right),
\end{eqnarray}
\end{widetext}
where a derivative on a internal state is used as a expression for  derivative on the position wave-function of that internal state. The first term Eq. \eqref{nonad1} is the adiabatic part of the kinetic operator. $K_{1}$ denotes the non-adiabatic contribution and couples the $\mathcal{N}=1$ states to $\mathcal{N}=0$ level. Such a transition has energy gap of $\hbar B_e$. In the present case we find that $||K_{ 1}||/(\hbar B_e) \sim 10^{-7} $ and as a result its effect can be neglected. The operator $K_2$ in Eq.~\eqref{nonad1} denotes non-adiabatic corrections leading to coupling between the trapped states. In the linear field regime, ($E_{-}\propto -ie^{i\varphi}$), this term can be shown to be proportional to $(\nabla\ket{-\varphi})\bra{-\varphi}$. Again as $||K_{2}||/K \sim 10^{-4} $, we can neglect its effect compared to the adiabatic contribution.

Similarly one can show that the non-adiabatic coupling between the other states also results in small corrections to the kinetic operator.

\section{Loss due to hyperfine structure induced coupling}
\label{app:hyperfine}
We give the detailed derivation of the hyperfine structure induced loss rates as presented in Section \ref{subs:hfloss}. The loss rate arises due to the off-diagonal elements of
$H_{01}, H_{10}$ in Eq.~\eqref{eigeneq22} which are of order $\delta\ll 1$ (cf. Table \ref{tab:table3}) and couple each trapped state to a continuum state with different hyperfine structure. We are interested in a regime where the energy of $j=0$ trapped state $2\omega[a_d,Z] - |E_0-E_1|<0$. Moreover, we are interested in a regime where the width of the trapped state is much smaller than the classical turning point radius. Away from the size of our square cell, $R>a_d/2$, we assume that the untrapped state is essentially a free particle with respective energy-independent two-dimensional density of states. Assuming such density of state is an approximation which can drastically change in presence of a resonance. In such cases, one can use the magnetic field to tune $|E_0-E_1|$ away from such resonance.

The relevant equations for the $j=0$ trapped state is derived from Eqs.~\eqref{eigeneq21} and ~\eqref{eigeneq22},
\begin{widetext}
\begin{eqnarray}\label{hypercont1}
\begin{bmatrix}
- \left(\partial^2_{\tilde{R}} + \frac{\partial_{\tilde{R}}}{\tilde{R}} - \frac{1}{\tilde{R}^2}\right) + \tilde{R}^2 - \frac{2E_0}{\hbar\omega[a_d,Z]}&  -\frac{3\delta\tilde{R}^2}{4}\\
-\frac{3\delta\tilde{R}^2}{4} &  - \left(\partial^2_{\tilde{R}} + \frac{\partial_{\tilde{R}}}{\tilde{R}} - \frac{5}{R^2}\right) + V[\tilde{R}] - \frac{2E_0}{\hbar\omega[a_d,Z]}
\end{bmatrix} &=& \frac{2\epsilon}{\hbar\omega[a_d,Z]} \begin{bmatrix}
{t}^0_1[\tilde{R}]\\
{u}^1_{-1}[\tilde{R}]
\end{bmatrix}, \nonumber\\
&&
\end{eqnarray}
\end{widetext}
We are interested in a region where $|E_0-E_1|> 2\omega[a_d,Z]$ {i.e.,} where resonant tunneling is possible when the energy of the continuum state is 
$\epsilon_{\rm res}=2\omega[a_d,Z] - |E_0-E_1|$. This corresponds to a
situation where the center of the trap $R=0$ is situated in a classically
forbidden region of the continuum state.  

We first numerically calculate the position dependence of $\ket{u^1,-1}$ with
energy $\epsilon_{\rm res}$ where we replace the shape of the potential for
the continuum state as: $V[\tilde{R}]=-\tilde{R}^2/2, \tilde{R}<\tilde{a}_d/2$
and $ V[\tilde{R}]=V[\tilde{a}_d/2]=-\tilde{a}^2_d/8, \tilde{R} \geq
\tilde{a}_d/2$. The reason behind this substitution is that the potential for
the continuum state has a downward curvature near the center of the trap with
the minimum residing at the boundary of the square cell. Our trapped states
are concentrated near the center of the square cell ($\sigma[a_d,Z]/a_d \ll
1$) and we are interested in classically forbidden energy regimes $\epsilon \sim V[a_d/2]$. With this substitution, we find the solution for $\ket{u^1,-1}$ inside a region of $\tilde{R} \in [0,\tilde{R}_{\rm max}]$. To find the density of states at the resonant energy, we first notice that as $\tilde{R}\to\infty$, the solution should approach the free-particle wave-function $\propto \mathcal{J}_{\sqrt{5}}[\alpha_{M}\tilde{R}/\tilde{R}_{\rm max}]$, where $\alpha_{M}$ is the $M$th
zero of the Bessel function $\mathcal{J}_{\sqrt{5}}[x]$.
Here, we numerically find $M_0$ by maximizing the overlap function: $\mathcal{O}[M] = |\braket{M\mid u^1,-1}|^2$ where
$\ket{M}=\sqrt{2}\tilde{R}^2_{\rm
  max}\mathcal{J}_{\sqrt{5}}[\alpha_{M}\tilde{R}/\tilde{R}_{\rm
  max}]/\mathcal{J}_{\sqrt{5}+1}[\alpha_{M}]$. The density of states is then given by $\mathcal{D} [M_0]=
\tilde{R}^2_{\rm max}/(2\pi\alpha_{M_0})$. Then using Fermi's golden rule, we express the transition rate from the lowest
energy trapped state {to the continuum} as,
\begin{equation}\label{decaystate0}
\frac{\gamma^0_{\rm hf}}{\hbar\omega[a_d,Z]} = \frac{9\pi\delta^2\mathcal{D}[M_0]}{32} |\braket{u^1,-1\mid\tilde{R}^2\mid t^0,01}|^2
\end{equation}

On the other hand, the transition rate from the trapped state $\ket{t^1,01}$ happens at a positive energy and is given by,
\begin{equation}\label{decaystate1}
\frac{\gamma^1_{\rm hf}}{\hbar\omega[a_d,Z]} = \frac{9\pi\delta^2\mathcal{D} [M_{\rm res}]}{32} |\braket{u^0,3\mid\tilde{R}^2\mid t^1,01}|^2,
\end{equation}
where the resonant condition now reads: $\alpha^2_{\ell,M_{\rm
    res}}/\tilde{R}^2_0=4+2|E_0-E_1|/(\hbar\omega[a_d,Z])+\tilde{R}^2/2$, cf. Sec.~\ref{subs:hfloss}.

\section{Loss rate due to the $R^3$-dependence of electric field} 
\label{app:r3loss}
\subsection{Loss rate for $0$D trap}
Including the correction due to $\propto R^3\exp[-3i\varphi]$ in Eq.~\eqref{series0d} and going to the transformed basis, the equation of motion for the $\ell=1$ trapped state Eq.~\eqref{trapdiag} reads
\begin{align}\label{eqhighersup}
\frac{-1}{2}& \left(\partial^2_{\tilde{R}} + \frac{\partial_{\tilde{R}}}{\tilde{R}} - \frac{1}{\tilde{R}^2}-2\tilde{R}^2\right)t^j_1[\tilde{R}]  +\frac{f_{-3}[a_d,Z]K}{f_{\perp}[a_d,Z]\omega[a_d,Z]} \nonumber\\
&\times \tilde{R}^4\sum_{\eta=\pm1}\left(t^i_{1+4\eta}[\tilde{R}]-\frac{3}{4}u^j_{1+4\eta}[\tilde{R}]\right) = \frac{\epsilon_{1}+E_i}{\hbar\omega[a_d,Z]}t^j_1[\tilde{R}],
\end{align}
where $\tilde{R}=R/\sigma[a_d,Z]$ and the last term on the right hand
side arises from the $R^3$ correction to the field. While deriving
Eq.~\eqref{eqhighersup}, we have neglected the effect of
$f_{1\perp}[a_d,Z]$ in Eq.~\eqref{series0d}, as we consider regions
with $N_{\rm max} > 1$ where its effect is negligible as it only
renormalizes the trapping frequency by a factor $
1+(f_{1\perp}[a_d,Z]/f_{\perp}[a_d,Z])^2 \approx 1$ (cf.\
Fig.~\ref{figure2a}). As a result we only consider the effect of
$f_{-3}[a_d,Z]$ components in Eq.~\eqref{series0d}.
Next, we solve Eq.~\eqref{eqhighersup}
perturbatively with the zeroth order solution given by
Eq.~\eqref{oscen}. The second last term in Eq.\eqref{eqhighersup} couples
$j,1,N$ to $\ket{t^1_{-3N'}},\ket{t^1_{5N'}}$ states. As from
Fig.~\ref{figure2a}, ${|f_{-3}[a_d,Z]}|/{f_{\perp}[a_d,Z]} \sim
10^{-3}$, we express the perturbed wavefunction as
\begin{eqnarray} \label{pert4sup}
\ket{\underline{t}^j, 1N}&=&\ket{t^j, 1N} + \frac{f_{-3}[a_d,Z] K}{f_{\perp}[a_d,Z]\hbar\omega[a_d,Z]} \sum_{\ell'=-3,5 } \nonumber\\
&\times& \sum_{N'\neq N}\frac{V_{N,\ell,N'}\ket{t^j, \ell'N'}}{2(N-N')+1-\sqrt{(\ell-1)^2+1}},  \nonumber\\
\end{eqnarray}
 where $V_{N;\ell,N'}=\braket{t^j, 1N\mid\tilde{R}^4
\mid t^j, \ell N'}$ is a
dimensionless number. These
  admixtures can have a strong impact
on the loss rate, especially 
for $\ell=1$, which is lossless to zeroth order as seen from
Eq.~\eqref{decay1} and Fig.~\ref{figure3}, but becomes lossy due
to the small admixture of the degenerate states $\ell=-3,5$. It is clear that the correction does not couple different hyperfine manifolds, so we will drop the hyperfine subscript $j$. From
Eq.~\eqref{pert4sup}, we find that the modified decay rate for the $N=0$ motional state of $\ell=1$ is given by Eq.~\eqref{decaytrapr4}.

The next source of loss originates from the last term in
Eq.~\eqref{eqhighersup} which couples the trapped $t_\ell$ state to the
continuum states $u_{\ell\pm 4}$. Using Fermi's golden rule, the loss rate for the $N=0,\ell=1$ state consequently is given by Eq.~\eqref{decaycontr4}.

\subsection{Loss rate for $1$D trap}
For $1$D trap, in the electric field expansion in Eq.~\eqref{fitting2}, both $\exp[\pm 1 3\phi_q]$ terms are present. As a result, the equivalent of Eq.~\eqref{eqhighersup}
\begin{align}\label{eqhighersup1d}
\frac{-1}{2}& \left(\partial^2_{\tilde{R}} + \frac{\partial_{\tilde{R}}}{\tilde{R}} - \frac{1}{\tilde{R}^2}-2\tilde{R}^2\right)t^j_1[\tilde{R}]  +\frac{K\tilde{R}^4}{2F_{\perp}[a_d,Z]\omega[a_d,Z]} \nonumber\\
&\times \sum_{\eta=\pm1}\left(F_{-3}[a_d,Z]t^j_{1+4\eta}[\tilde{R}]+F_{3}[a_d,Z]t^j_{1+2\eta}[\tilde{R}] \right. \nonumber\\
&+\left. \frac{3}{4}F_{-3}[a_d,Z]u^j_{1+4\eta}[\tilde{R}]+\frac{3}{4}F_{3}[a_d,Z]u^j_{1+2\eta}[\tilde{R}]\right) \nonumber\\
&= \frac{\epsilon_{1}+E_i}{\hbar\omega[a_d,Z]}t^j_1[\tilde{R}].
\end{align}

Then we get the loss rates for $1$D following the same procedure as the previous subsection.

\section{Total molecular loss for a different magnetic field}
\label{app: hyperfinepic}

\begin{figure}
    \includegraphics[width=.5\textwidth]{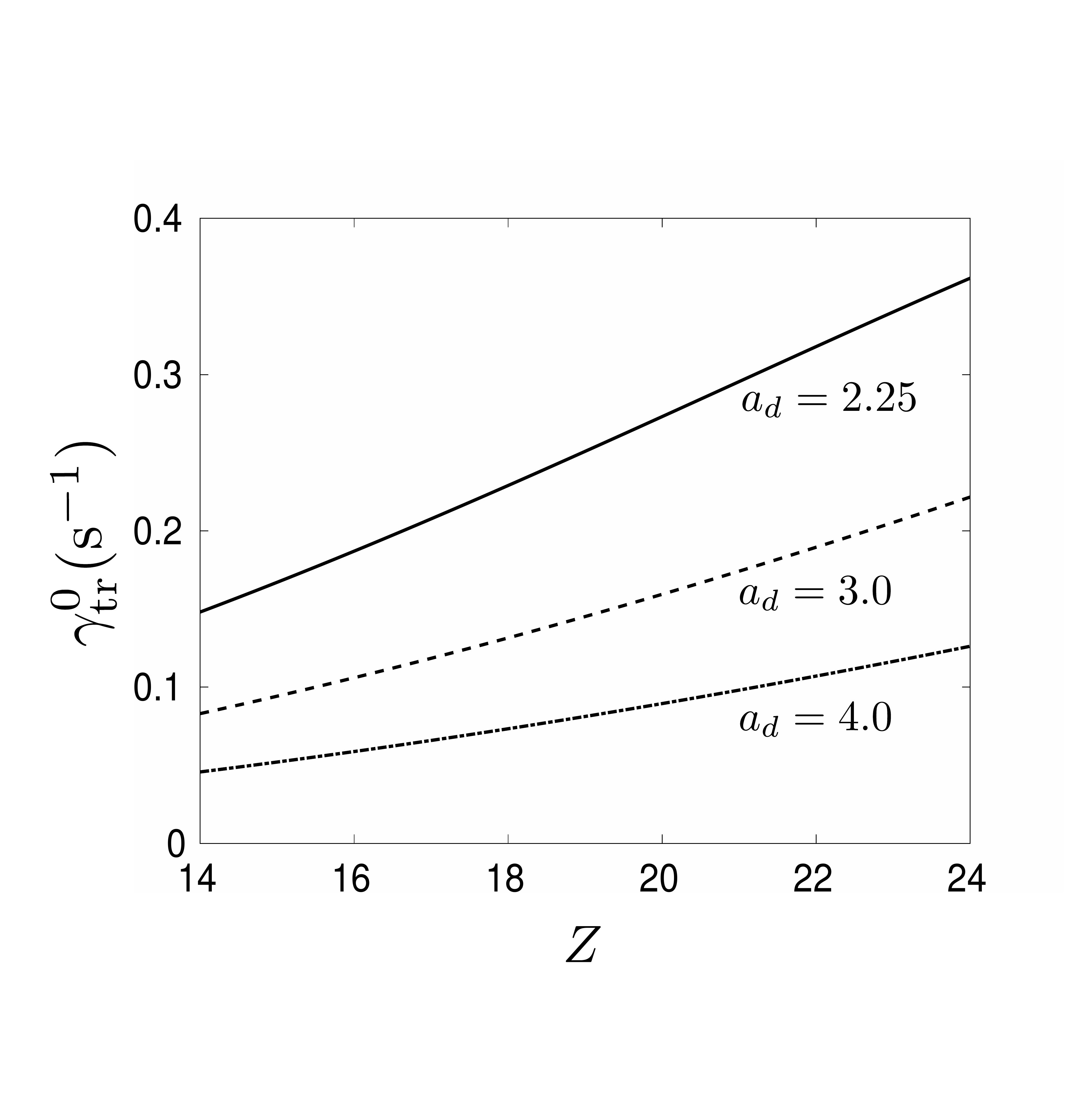}
    \caption{Total non-adiabatic and hyperfine-induced loss rate $\gamma^0_{\rm tr}$ is plotted as a function of $Z$ for $a_d=2.25$ (solid line), $3.0$ (dashed line), $4.0$ (dash-dotted line).} 
        \label{supfigure1}
\end{figure}

Here we show the total molecular loss rate $\gamma_{\rm tr}$ as defined in
Eq.~\eqref{eq:10} for a magnetic field $B_0=5$T. For such a magnetic field the
hyperfine parameters of Eqs.~(\ref{magstates0}, \ref{magstates1}) for the
$j=0,1$ states are given by $\delta \approx 0.01$ and $|E_1-E_0|/(\hbar B_e)
\approx 0.027$. For such parameters, along with the ferroelectric parameter of
the $0$D trap from Table~\ref{tab:table1},
we plot the loss rate for different values of $a_d$ in Fig.~\ref{supfigure1}.
%---------------------------
%----------------------------

\section{Effect of $E_Z$ field}
\label{app:ezfield1}
An additional loss channel arises due to the second-order Stark term in Eq.~\eqref{effpot1},
\begin{align}\label{zloss}
V^{\mathcal{N}=1}_{Z}&= -\frac{E^2_Z[\vec{\rho}] a^2_{\rm mf}}{20\hbar B_e} \sum_{\substack{\mathcal{M}_1=\pm 1 \\ \mathcal{I}_{\rm col}}} \ket{1,\mathcal{M}_1,\mathcal{I}_{\rm col}}\bra{1,\mathcal{M}_1,\mathcal{I}_{\rm col}}.
\end{align}
By using the fitting potential of Eq.~\eqref{series0d}, $E_Z[\vec{\rho}]\approx f_z[a_d,Z]R^2\sin[2\varphi], R\rightarrow 0$ and going to the position basis as used
for Eq.~\eqref{trapdiag}, the transformed equation reads,
\begin{widetext}
\begin{equation}\label{eqZ}
 \left[-\partial^2_{\tilde{R}} - \frac{\partial_{\tilde{R}}}{\tilde{R}} + \frac{(\ell-1)^2+1}{\tilde{R}^2}
+\tilde{R}^2 - \frac{f^2_z[a_d,Z]K}{2f^2_{\perp}[a_d,Z]\hbar\omega[a_d,Z]}\tilde{R}^4\right] 
 t^j_\ell[\tilde{R}] +\frac{f^2_z[a_d,Z]K}{4f^2_{\perp}[a_d,Z]\hbar\omega[a_d,Z]}\tilde{R}^4\left(t^j_{\ell+4}[\tilde{R}]+t^j_{\ell-4}[\tilde{R}]\right) = \frac{2(\epsilon_{\ell}+E_j)}{\hbar\omega[a_d,Z]}t^j_\ell[\tilde{R}]. 
\end{equation}
\end{widetext}
We solve Eq.~\eqref{eqZ} perturbatively with the zeroth order solution
given by Eq.~\eqref{oscen}. The attractive $R^4$ will lower the
barrier due to the quadratic potential for large $R$ and will lead to
tunneling loss. From a semi-classical WKB approximation, the tunneling
loss rate is found to be negligible compared to the other time scales involved
in our system and as a result we neglect this effect. The last two terms will
perturbatively couple $t^j_1$ to $t^j_{-3,5}$. Again, there is no
coupling between different hyperfine $j$ states and as a result we drop the label for rest of the section. The perturbed state is given by,
\begin{eqnarray}
\ket{\underline{t}^j, 1N}&=&\ket{t^j, 1N} + \frac{f^2_{z}[a_d,Z] K}{4f^2_{\perp}[a_d,Z]\hbar\omega[a_d,Z]}\sum_{\ell'=-3,5 } \nonumber\\
&\times& \sum_{N'\neq N}\frac{V_{N,\ell,N'}\ket{t^j, \ell'N'}}{2(N-N')+1-\sqrt{(\ell-1)^2+1}},  \nonumber\\
\end{eqnarray}
where $V_{N;\ell,N'}$ is defined below Eq.~\eqref{pert4sup}. Such a
superposition will have an impact on the loss rate, specially
for $\ell=1$. In the zeroth order, the $\ell=1$ is the only loss-less
state as seen from Eq.~\eqref{decay1} and Fig.~\ref{figure3}. The loss rate is same as Eq.~\eqref{decaytrapr4} with $\frac{f_{-3}[a_d,Z]}{f_{\perp}[a_d,Z]}$ replaced by $(\frac{f_{z}[a_d,Z]}{2f_{\perp}[a_d,Z]})^2$. We find that $\frac{f_{z}[a_d,Z]}{f_{\perp}[a_d,Z]} \sim 10^{-1}$ and the loss rate is $\sim 10^{-2}$s$^{-1}$ for $a_d=2.25,Z=16$.  

\section{Effect of coupling to $\ket{\mathcal{N}=1,\mathcal{M_N}=0}$ state}
\label{app:ezfield2}
One of the loss channels we have neglected so far is due to the coupling between a $\ket{\pm 1}=\ket{\mathcal{N}=1,\mathcal{M_N}=\pm 1}$ and  $\ket{0}=\ket{\mathcal{N}=1,\mathcal{M_N}=0}$ states. Such coupling arises again via quadratic Stark shift and the corresponding coupling Hamiltonian is given by,
\begin{equation} \label{lossZ}
V^{\mathcal{N}}_{\rm 0\perp} = \frac{a^2_{\rm mf}E_Z}{\sqrt{2}\hbar B_e}\left(E_{+}V^{\mathcal{N}}_{0+}+E_{-}V^{\mathcal{N}}_{0-}\right),
\end{equation}
where
\begin{eqnarray}
V^{\mathcal{N}}_{0+}&= &\sum_{\epsilon=0,1}\sum_{\mathcal{N}',\mathcal{M_{N'}}}\frac{\bm{T^1_\epsilon}\ket{\mathcal{N}',\mathcal{M_{N'}}}\bra{\mathcal{N}',\mathcal{M_{N'}}}\bm{T^1_{\epsilon}}}{E_\mathcal{N}-E_\mathcal{N'}},\\
V^{\mathcal{N}}_{0-}&=&\sum_{\epsilon=0,-1}\sum_{\mathcal{N}',\mathcal{M_{N'}}}\frac{\bm{T^1_\epsilon}\ket{\mathcal{N}',\mathcal{M_{N'}}}\bra{\mathcal{N}',\mathcal{M_{N'}}}\bm{T^1_{\epsilon}}}{E_\mathcal{N}-E_\mathcal{N'}},
\end{eqnarray}

For the first rotational level $\mathcal{N}=1$, the internal states characterized by: $\ket{-1}=\ket{\mathcal{N}=1,\mathcal{M_N}=-1} , \ket{0}=\ket{\mathcal{N}=1,\mathcal{M_N}=0}, \ket{1}=\ket{\mathcal{N}=1,\mathcal{M_N}=1}$. With these basis states, the various components of the coupling matrices are given by, \[ V^{1}_{0+} \approx
\begin{bmatrix}
0 & 3/20 & 0 \\
0 & 0 & 3/20 \\
0 & 0 & 0
\end{bmatrix} \\,
V^{1}_{0-} \approx
\begin{bmatrix}
0 & 0 & 0 \\
3/20 & 0 & 0 \\
0 & 3/20 & 0
\end{bmatrix}.
\]

By including the energy shift of the $\ket{0}$ state due to laser potential,
$E_0=-\alpha_0I_0$ \cite{jin} and using Fermi's Golden Rule, the loss rate is
proportional to the coupling constant, $
|\bra{t;\ell;N;\mathcal{N}=1}V^{\mathcal{N}}_{\rm 0\perp}\ket{0,\vec{k}}|^2$
where $\ket{0,\vec{k}}=e^{i \vec{k}\cdot\vec{R}}\ket{0}/\sqrt{A}$ is a 2D free
particle state with momentum, $\vec{k}$,
$|\vec{k}|=\sqrt{(\epsilon_{\ell,N}[Z]+E_0)/K}$. We find that specifically for the $\ell=1$ state, i.e., the state $\ket{t;\ell=1;N;\mathcal{N}=1}$, the above integral vanishes, i.e., $|\bra{t;\ell=1;N;\mathcal{N}=1}V^{\mathcal{N}}_{\rm 0\perp}\ket{0,k}|=0$ as $E_Z \propto \sin 2\varphi$ (Fig.~\ref{figure1}c).

\section{Effect of Casimir-Polder force}  \label{app:casimir}
Another possible modification arise from attractive Casimir-Polder
potential. Using similar effective medium arguments, we can infer that the
important contribution comes from the substrate. For a planar substrate in the
non-retarded regime, from \cite{caspol2} we can write down the Casimir-Polder potential for the level $\ket{\mathcal{N}=1,\mathcal{M_N}=\pm1}$ as,
$$
V_{\rm Casimir}=-\frac{3\mu^2}{160\pi\epsilon_0}\frac{\epsilon_{\rm s}-1}{\epsilon_{\rm s}+1}\frac{1}{Z^3},
$$
where $Z$ is the distance (not scaled with $r_d$), $\epsilon_{\rm s}$ is the static dielectric constant of the substrate ($\epsilon_{\rm s}\gg 1$). As we trap our molecule at a distance of $Z_0\approx 420$nm from the substrate, strength of the Casimir-Polder energy for RbCs molecule is $V_{\rm Casimir}/\hbar B_e \approx 10^{-7}\ll \hbar\omega[Z_0]$, much weaker than the trapping potential.

\section{Expansion of $E_{-}[m_x,m_y;\vec{R},Z]$ in Eq.~\eqref{efield} and the resulting total azimuthal field $E_{-}[\vec{\rho}]$}
\label{1dapp}

To look for the behaviour of the electric field, we first chose a rectangular configuration from Fig.~\ref{figure1}(b) with the nano-rod positions, $(m_x(qa_{\rm latt}-a_d),m_ya_d)$ and for a fixed $q$ and $m_x=\pm 1, m_y=\pm1$. The electric field for such a configuration can be written as 
\begin{align}\label{E_field_supp}
E_{-}[q,\bm{m};\vec{\rho}] &= -  (-1)^{\frac{m_x+m_y}{2}} e^{-i\tilde{\phi}_{q,\bm{m}}} \int d\vec{r} \left(E[Z+h]-E[Z]\right) \nonumber\\
& \times 
\left({R}_{q,\bm{m}}-r e^{-i(\phi-{\phi}_{q,\bm{m}})}\right),  \nonumber\\
E[Z] & =\left(Z^2+|\vec{R}_{q,\vec{m}}-\vec{r}|^2\right)^{-3/2}
\end{align}
where $\vec{R}_{q,\bm{m}}=\vec{R}-\vec{F}_{q,\bm{m}}$, $\bm{m} \equiv (m_x,m_y)$ and
{$\tan{\phi}_{q,\bm{m}}=(Y-m_ya_d)/(X-m_x(qa_{\rm
    latt}-a_d))$}.
The total
field components are given as the sum of the contributions of all
nano-rods by  $E_{-q}[\vec{\rho}]$:
\begin{align}\label{sumefield}
  E_{-q}[\vec{\rho}]&=\sum_{\bm{m}}E_{-}[q,\bm{m};\vec{\rho}].
\end{align}

{\bf Terms of order $R$:} As we are interested in the field near the center of the rectangle,
we carry out the integration over $r$ in Eq.~\eqref{E_field_supp} and we expand the resulting expression as a function of $R \ll a_d/2$. The electric field can be written as,
\begin{align}\label{expandefield}
E_{-}[q,\bm{m};\vec{\rho}] & =(-1)^{\frac{m_x+m_y}{2}}s_{m_x,m_y}e^{-i\tilde{s}_{m_x,m_y}\phi_0} \nonumber\\
& \times \left(1+i\{m_xa_xY-m_ya_dX\} + \mathcal{O}[R^2_q]\right)\nonumber\\
& \times \left(g_0[R_q,Z]+g_1[R_q,Z]\{m_xa_xX+m_ya_dY\} + \mathcal{O}[R^2_q,Z]\right),
\end{align} 
where we have defined $a_x=qa_{\rm latt}-a_d$ and $R^2_q=a^2_x+a^2_d$ and
$g_{0,1}[R_q]$ are functions resulting from the integration which only depends
on $R_q$ and $Z$. Moreover we define the angle {$\tan \phi_0 = (a_d/a_x)$}. The sign functions are defined as: ${s}_{1,1}=s_{1,-1}=1; s_{-1,1}=s_{-1,-1}=-1$ and $\tilde{s}_{1,1}=\tilde{s}_{-1,-1}=1; \tilde{s}_{-1,1}=\tilde{s}_{1,-1}=-1$. 
Now from Eqs.~\eqref{expandefield} and Eq.~(\ref{sumefield}), by carrying out the summation for terms $\propto R$ we find that, 
$ E_{-q}[\vec{\rho}] \propto R e^{i\phi}$. As a result, to get the full electric field one sums over terms similar to Eq.~\ref{sumefield} over different $q$ which only changes the strength of the leading order term.

\section{Derivation of simulation Hamiltonian}
\label{app:hsim}
Here we derive the equation in Eq.~\eqref{hsim}. At first we only consider a single molecule near the ferroelectric substrate in a linearly $Z$-polarized
microwave field $\vec{E}_{\rm mw}=E_0 \cos{\Omega t}\hat{Z}$ where
$\Omega=4B_e+\Delta$ with $\Delta \ll B_e$. Such a frequency will resonantly couple $\ket{\mathcal{N}=1}$ to $\ket{\mathcal{N}=2}$ state. Our Hamiltonian is then given by,
\begin{widetext}
\begin{eqnarray}\label{fullham1}
H &=& \sum_{\mathcal{N},\mathcal{M_N}}E_{\mathcal{N}} \ket{\mathcal{N},\mathcal{M_N}}\bra{\mathcal{N},\mathcal{M_N}} + 
\sum_{\substack{|\mathcal{N}-\mathcal{N'}|=1, \\\mathcal{M_N},\mathcal{M_N'}}}H_{\rm mf}[\mathcal{N'},\mathcal{M_N'};\mathcal{N},\mathcal{M_N}] \ket{\mathcal{N'},\mathcal{M_N'}}\bra{\mathcal{N},\mathcal{M_N}} \nonumber\\
 &+& \cos{\Omega t}\sum_{\substack{|\mathcal{N}-\mathcal{N'}|=1, \\\mathcal{M_N}}}g_{\rm mw}[\mathcal{N'},\mathcal{N},\mathcal{M_N}] \ket{\mathcal{N'},\mathcal{M_N}}\bra{\mathcal{N},\mathcal{M_N}},
\end{eqnarray}
%\end{widetext}
where the molecular rotational energy, $E_{\mathcal{N}}=\hbar B_e \mathcal{N}(\mathcal{N}+1)$, the molecule-ferroelectric coupling, 
$H_{\rm mf}[\mathcal{N'},\mathcal{M_N'};\mathcal{N},\mathcal{M_N}]=\bra{\mathcal{N'},\mathcal{M_N'}} H_{\rm mf}\ket{\mathcal{N},\mathcal{M_N}}$ with $H_{\rm mf}$ is defined in Eq.~\eqref{molferro} and is position dependent.
The microwave coupling is given by, $g_{\rm
  mw}[\mathcal{N'},\mathcal{N},\mathcal{M_N}]=\mu E_0
\bra{\mathcal{N'},\mathcal{M_N}}
\bm{T^1_0}\ket{\mathcal{N},\mathcal{M_N}}$. We apply the unitary
transformation, $\bm{U}_t=\exp\left[-i
  \sum_{\mathcal{N},\mathcal{M_N}}E_{\mathcal{N}}
  \ket{\mathcal{N},\mathcal{M_N}}\bra{\mathcal{N},\mathcal{M_N}} t\right]$ and
carry out the transformation $H'=\bm{U}^{\dagger_t} H \bm{U}_t - i
\bm{U}^{\dagger}_t [d_t\bm{U}_t] $. As a result, Eq.~\eqref{fullham1} becomes,
\begin{eqnarray}\label{fullham2}
H' &=&  \Delta \sum_{\mathcal{M}_2} \ket{2,\mathcal{M}_2}\bra{2,\mathcal{M}_2} +  \frac{1}{2}\sum_{\mathcal{M}_1}(g_{\rm mw}[2,1,\mathcal{M}_1] \ket{2,\mathcal{M}_1}\bra{1,\mathcal{M}_1}+g_{\rm mw}[1,2,\mathcal{M}_1]\ket{1,\mathcal{M}_1}\bra{2,\mathcal{M}_1}) \nonumber\\
&+&\sum_{\substack{|\mathcal{N}-\mathcal{N'}|=1, \\\mathcal{M_N},\mathcal{M_N'}}}H_{\rm mf}[\mathcal{N'},\mathcal{M_N'};\mathcal{N},\mathcal{M_N}]\exp\left[i (E_{\mathcal{N'}}-E_{\mathcal{N}})t \right] \ket{\mathcal{N'},\mathcal{M_N'}}\bra{\mathcal{N},\mathcal{M_N}} \nonumber\\
 &+& \frac{1}{2}\sum_{\mathcal{M}_1}(g_{\rm mw}[2,1,\mathcal{M}_1]e^{i8B_et} \ket{2,\mathcal{M}_1}\bra{1,\mathcal{M}_1}+g_{\rm mw}[1,2,\mathcal{M}_1]e^{-i8B_et}\ket{1,\mathcal{M}_1}\bra{2,\mathcal{M}_1})\nonumber\\
 &+&\cos{\Omega t}\sideset{}{'}\sum_{\substack{|\mathcal{N}-\mathcal{N'}|=1, \\\mathcal{M_N}}}g_{\rm mw}[\mathcal{N'},\mathcal{N},\mathcal{M_N}] \exp\left[i (E_{\mathcal{N'}}-E_{\mathcal{N}})t \right] \ket{\mathcal{N'},\mathcal{M_N}}\bra{\mathcal{N},\mathcal{M_N}},
\end{eqnarray}
\end{widetext}
where in the first line we have defined the time-independent component
in the transformed Hamiltonian and in the last line, the summation
$\sideset{}{'}\sum$ excludes transitions between $\mathcal{N}=1$ and
$\mathcal{N}=2$ states. We carry out a Floquet-Magnus expansion
\cite{floq} up to second order ($1/\Omega^2$) which results in the effective Hamiltonian in the $\mathcal{N}=1,2$ subspace
\begin{widetext}
\begin{eqnarray}\label{floqham}
H_{\rm eff}&=&\Delta \sum_{\mathcal{M}=-2}^{2}
\ket{2,\mathcal{M}}\bra{2,\mathcal{M}} + g\sum_{\mathcal{M}=\pm 1}
(\ket{2,\mathcal{M}}\bra{1,\mathcal{M}}+h.c.)+g\sqrt{2/3}
(\ket{2,0}\bra{1,0}+\mathrm{h.c.})
\nonumber\\
&+&\sum_{\mathcal{M'}_1,\mathcal{M}_1}V^{1}[\mathcal{M'}_1,\mathcal{M}_1]\ket{1,\mathcal{M'}_1}\bra{1,\mathcal{M}_1} 
+\sum_{\mathcal{M}_2,\mathcal{M'}_2}V^{2}[\mathcal{M'}_2,\mathcal{M}_2]\ket{2,\mathcal{M'}_2}\bra{2,\mathcal{M}_2}, 
\end{eqnarray}
\end{widetext}
where the matrix elements in $V^{\mathcal{N}}$ is given by
Eq.~\eqref{effpot1} and $g=\mu E_0/\sqrt{20}$. The first line
represents the time-independent part and the second line comes from
the first term in the Magnus expansion \cite{floq}. As we noticed from
our discussion of Eqs.~\eqref{effpot1}, the terms of the last line
will give rise to trapped states $\ket{t;\mathcal{N}=1}$ and
$\ket{t;\mathcal{N}=2}$ with energy $\hbar \omega[Z_1]$ and $\hbar
\omega[Z_2]$. Then, if we project the Hamiltonian in
Eq.~\eqref{floqham} to the motional ground states in the two
  $\mathcal{N}$ subspaces and define the operators
$\bm{S}^{+}=\ket{t;\mathcal{N}=1}\bra{t;\mathcal{N}=1},
\bm{S}^{z}=\ket{t;\mathcal{N}=2}\bra{t;\mathcal{N}=2}-\ket{t;\mathcal{N}=1}\bra{t;\mathcal{N}=1}$,
and $\bm{S}^{-}=[\bm{S}^{+}]^{\dagger}$, this gives rise to the last
two terms in Eq.~\eqref{hsim} where we have introduced an additional
subscript $q$ to denote the mean position of the trapped
state. Next we look into the situation when each sites in the periodic
potential is filled with one molecule and consider the effects of
the dipolar interaction. Dipolar interaction between the two molecules is given by,
\begin{equation}
V_{\rm dip} = \frac{\bm{\mu}_1\cdot\bm{\mu}_2 - 3(\bm{\mu_1}\cdot\hat{\rho}_{12})(\bm{\mu_2}\cdot\hat{\rho}_{12})}{|\vec{\rho}_{12}|^3},
\end{equation}
where $\bm{\mu}_{1,2}$ are the dipole moment operator for molecules are position $\vec{\rho}_{1,2}$ and $\vec{\rho}_{12}=\vec{\rho}_{1}-\vec{\rho}_{2}$. Between molecules at site $q$ and $q'$, the resonant dipolar interaction is given by ,
\begin{align}
& \bra{t; \mathcal{N}=2}_q\bra{t; \mathcal{N}=1}_{q'}V_{\rm dip}\ket{t;\mathcal{N}=1}_q\ket{t;\mathcal{N}=2}_{q'}   \nonumber\\ & = \sum_{\mathcal{M_N}=\pm 1}\frac{\mu^2|\bra{\mathcal{N}=2,\mathcal{M_N}}\bm{T^1_0}\ket{\mathcal{N}=1,\mathcal{M_N}}|^2}{16\pi\epsilon_0|q-q'|^3a^3_d} 
\end{align}
where we assume that $1 \gg \sigma[a_d,Z]/a_d$ and give
rise to the first term in Eq.~\eqref{hsim}. 
There will also be  losses due to dipolar coupling of untrapped state. After some algebra, one such lossy term can be written as 
\begin{widetext}
\begin{align}
 & \bra{\mathcal{N}=2,\mathcal{M_N}=0;k}\bra{\mathcal{N}=1,\mathcal{M_N}=0;k'}V_{\rm dip}\ket{t;\mathcal{N}=1}_q\ket{t;\mathcal{N}=2,-}_{q'}   \approx  -\frac{\mu^2|\exp[-\tilde{R}^2/2-i\varphi-i\vec{k}\cdot\vec{\tilde{R}}]d^2\tilde{R}|^2}{8\pi\epsilon_0|q-q'|^3a^3_d} \nonumber\\ & \times \bra{\mathcal{N}=2,\mathcal{M_N}=0}\bm{T^1_{-1}}\ket{\mathcal{N}=1,\mathcal{M_N}=1}\bra{\mathcal{N}=1,\mathcal{M_N}=0}\bm{T^1_{-1}}\ket{\mathcal{N}=2,\mathcal{M_N}=1} \nonumber\\
 &\propto \exp[-k^2/4] = \exp[-\frac{\alpha_0I_0}{4K}]
\end{align}
where we have used for the untrapped state $\bra{\mathcal{N}=2,\mathcal{M_N}=0;k}=\bra{\mathcal{N}=2,\mathcal{M_N}=0}\exp[-i\vec{k}\cdot\vec{\tilde{R}}]$ and $1 \gg \sigma[a_d,Z]/a_d$. For the last line we have used the fact that due to laser light, the $\mathcal{M_N}=0$ state is shifted by $\alpha_0 I_0$ (see Sec.~\ref{lassec}), the resonant condition reads $k^2=\alpha_0 I_0/K$. As a result the loss rate will be $\propto \exp[-\frac{\alpha_0I_0}{2K}]$.
\end{widetext}

\bibliographystyle{apsrev4-1}
\bibliography{ferromol}

\end{document}